\documentclass[%
 reprint,
 amsmath,amssymb,
 aps,
floatfix,
]{revtex4-2}

\usepackage{xspace}
\usepackage{subcaption}
\usepackage{multirow}
\usepackage{graphicx}
\usepackage{dcolumn}
\usepackage{bm}
\usepackage{hyperref}
\hypersetup{
	colorlinks=true,        
	linkcolor=blue,         
	citecolor=blue,        
	urlcolor=blue,           
}
\usepackage{dsfont}
\usepackage{comment}
\usepackage{color}
\usepackage[normalem]{ulem}
\usepackage{soul}
\usepackage{float}

\begin{document}

\preprint{APS/123-QED}

\title{A Possible Triple Formation Scenario of Binary Black Hole Merge  With One In Pair-instability Supernova Mass Gap}

\author{Tian Huang$^{1}$}
\author{Xizhen Lu$^{1}$}
\author{Guoliang L\"{u}$^{1}$}%
\email{guolianglv@sina.com}
\author{Chunhua Zhu$^{1}$}
\email{chunhuazhu@sina.cn}
\author{Sufen Guo$^{1}$}
\email{guosufen@xju.edu.cn}
\author{Helei Liu$^{1}$}
\author{Zhuowen Li$^{1}$}
\author{Zhijun Wang$^{1}$}
\author{Lei Li$^{2}$}
\author{Wei-Min Gu$^{3}$}
\author{Nurzada Beissen$^{4}$}

\affiliation{%
 $^{1}$School of Physical Science and Technology, Xinjiang University, Urumqi, 830046, China\\ 
 $^{2}$College of Physical Science and Technology, Yili Normal University, Yining ,Xinjiang, 835000, China\\ 
 $^{3}$Department of Astronomy, Xiamen University, Xiamen, Fujian 361005, P. R. China\\ 
 $^{4}$Institute for Experimental and Theoretical Physics, Al-Farabi Kazakh National University, Almaty 050040, Kazakhstan\\
}

\date{\today}

\begin{abstract}
Observations of binary black hole (BBH) mergers detected by LIGO---such as GW170729, GW190620, GW190706, GW230107, GW230820, and GW230928---feature high effective spins and primary black holes that fall squarely into the pair-instability supernova (PISN) mass gap ($\sim 45-130 \, M_{\odot}$). These events pose a significant challenge to standard stellar and binary evolution theories. To address this, we propose an isolated hierarchical triple stellar evolution channel. In this framework, tidal synchronization in tight inner binaries drives chemically homogeneous evolution (CHE), entirely bypassing giant expansion. A subsequent triple common envelope (TCE) evolution, triggered by the tertiary companion, rapidly drives the inner BBH to coalescence. Our model can provide a detailed evolutionary pathway that elegantly reproduces the properties of these GWs, such as GW190706. Assuming a low-metallicity environment ($Z = 0.001$), our framework predicts a volumetric merger rate of approximately $0.011 \, \mathrm{Gpc}^{-3}\mathrm{yr}^{-1}$ at $z \approx 0.68$, accounting for $22\%$ of the empirical rate for this mass regime in the GWTC-4 catalog. This study demonstrates that primordial triple interactions are a highly efficient avenue for populating the PISN mass gap.
\end{abstract}

\maketitle

\section{Introduction}

According to standard stellar evolution theory, the upper black hole mass gap between roughly $45 \, M_{\odot}$ and $130 \, M_{\odot}$ is a direct consequence of pair-instability supernovae (PISNe) \cite{2002ApJ...567..532H, 2019ApJ...887...53F}. In the cores of very massive stars, electron-positron pair production reduces radiation pressure, triggering explosive oxygen burning. Depending on core mass, this leads either to pulsational pair-instability supernovae (PPISNe)—where violent mass loss precedes black hole formation—or to full PISN. In the PISN case, the star undergoes complete thermonuclear disruption, leaving no compact remnant. The absence of remnants in this mass regime naturally produces the observed gap in the black hole mass distribution.

Recent breakthroughs in Galactic archaeology have provided direct observational confirmation for the physical reality of the PISN mechanism. The predicted nucleosynthetic fingerprints of Population III PISN—specifically a pronounced odd-even effect and extreme deficiency in odd-Z elements like sodium—have been extensively mapped by theoretical yield models \cite{2002ApJ...567..532H, 2002ApJ...565..385U, 2018ApJ...857..111T}. Building upon earlier candidate searches in the Galactic halo \cite{2014Sci...345..912A}, the extremely metal-poor star LAMOST J1010+2358 was recently found to exquisitely match the yields of a $260 M_{\odot}$ PISNe \cite{2023Natur.618..712X}. This observationally confirms that very massive stars capable of triggering PISNe indeed existed in the early Universe, perfectly aligning with cosmological simulations of Population III stars \cite{2002Sci...295...93A, 2004ARA&A..42...79B, 2014ApJ...781...60H}. 

Since gravitational waves (GWs) were first detected in 2015 \cite{2016PhRvL.116f1102A}, GWs astronomy has flourished. The LIGO-Virgo-KAGRA (LVK) collaboration has cataloged roughly several hundred binary black hole (BBH) mergers, yielding an unprecedented census of the cosmic black hole mass distribution \cite{2023PhRvX..13d1039A, 2025arXiv250818082T}.

However, GW events like GW190521($m_1 = 85^{+21}_{-14} \, M_{\odot}$) feature primary black holes squarely within this mass gap \cite{2020PhRvL.125j1102A}, posing a severe challenge to standard isolated binary evolution models and current mass loss prescriptions \cite{2016A&A...594A..97B, 2017ApJ...836..244W}. Consequently, the existence of mass gap black holes cannot be explained by dismissing the physical reality of PISNe; instead, it strongly demands novel astrophysical pathways capable of bypassing this mechanism.

Various physical channels are proposed to explain these mass gap BBHs, primarily including isolated binary evolution \cite{2019ApJ...887...53F, 2020A&A...636A.104B, 2021MNRAS.501.4514C}, stellar collisions \cite{ 2020A&A...640A..56R,2020ApJ...903...45K, 2020MNRAS.497.1043D}, dense star clusters \cite{2019PhRvD.100d3027R, 2019MNRAS.486.5008A}, and active galactic nucleus (AGN) disks \cite{2019PhRvL.123r1101Y, 2019ApJ...878...85S, 2021NatAs...5..749G}.

Classical isolated binary systems generally fail to evolve into black holes within the PISN mass gap. Although extreme super-Eddington accretion can theoretically allow a stellar component to grow into this forbidden mass regime \cite{2016Natur.534..512B}, population synthesis indicates that such gap pollution remains negligible \cite{2020ApJ...897..100V}. Moreover, the severe mass transfer required to achieve such extreme accretion inevitably widens the binary orbit, thereby preventing the resulting black holes from merging within a Hubble time \cite{2016A&A...588A..50M}. Consequently, the classical isolated binary channel is highly unlikely to produce observable mass gap mergers \cite{2021MNRAS.504..146V}.

Alternatively, three-dimensional simulations show head-on massive stellar collisions can form remnants that directly collapse into gap black holes, bypassing PISNe disruption \cite{2023MNRAS.519.5191B,Patton_2025}. However, the outcome depends heavily on envelope ejection efficiency \cite{1988ApJ...329..764L,2022MNRAS.511.5797M}, and the resulting black hole still requires subsequent dynamical capture to merge \cite{2014MNRAS.441.3703Z}.

Dense star clusters drive hierarchical mergers \cite{2019MNRAS.487.2947D, 2021NatAs...5..749G}. Deep gravitational potential wells retain a fraction of first-generation merger remnants \cite{2018PhRvL.121p1103F,2000ApJ...528L..17P, 2020ApJ...893...35D}. Through multi-body interactions, these remnants dynamically capture new companions, producing sequential mergers that comfortably surpass the PISNe limit \cite{2020ApJ...903...45K, 2021MNRAS.505..339M}.

The AGN disk channel traps stellar-mass black holes in dense gas environments \cite{2017ApJ...835..165B, 2017MNRAS.464..946S}. Gas drag induces orbital decay, driving bodies into deeper gravitational potential wells and thereby accelerating mergers \cite{2016ApJ...819L..17B}. High escape velocities prevent natal kick ejection, facilitating frequent multi-generation mergers that rapidly populate the mass gap \cite{2018ApJ...866...66M, 2020ApJ...898...25T,2019PhRvL.123r1101Y, 2021ApJ...908..194T}.

While the aforementioned dynamical channels successfully produce mass gap black holes, they strictly require exceptionally dense and specialized stellar environments. Conversely, classical isolated binary evolution struggles to populate this gap \cite{2016A&A...594A..97B, 2017MNRAS.470.4739S, 2019ApJ...882...36M}. Recent extensive observational surveys have firmly established that the vast majority of massive stars are not solitary; rather, they are born in multiple systems, with a remarkably high fraction of O-type stars—specifically $73\% \pm 16\%$—residing in hierarchical triple or higher-order architectures \cite{2012Sci...337..444S, 2013ARA&A..51..269D, 2014ApJS..215...15S, 2014ApJS..213...34K, 2017ApJS..230...15M}. Driven by these overwhelming observational facts, the hierarchical triple channel provides a highly competitive physical framework for populating the PISN mass gap purely in isolated galactic fields \cite{2017ApJ...841...77A, 2018ApJ...863...68L, 2019MNRAS.486.4443F, 2021ApJ...907L..19V, 2022A&A...661A..61T}.

In this work, we propose that hierarchical triple evolution offers a viable pathway to produce GW events with primary black holes in the PISN mass gap. We integrate triple common envelope (TCE) phases and chemically homogeneous evolution (CHE) into the triple-star evolution  code, \texttt{TSE},  to systematically investigate the formation of these asymmetric BBH mergers. The paper is organized as follows: Section \ref{sec:methods} introduces the methodology. Section \ref{sec:results} presents the initial parameters and simulation results, and Section \ref{sec:conclusion} provides our summary and conclusions.

\section{Methods}\label{sec:methods}

To accurately track the physical pathways leading to the formation of mass gap black holes, we employ the open-source numerical framework \texttt{TSE} \cite{2022MNRAS.516.1406S} to conduct forward simulations of the entire lifecycle of massive hierarchical triple systems. \texttt{TSE} deeply integrates the \texttt{MOBSE} module, which precisely treats extreme processes such as metallicity-dependent stellar wind mass loss and PISNe. Driven by stellar winds, the mass loss redistributes the orbital angular momentum, leading to adiabatic orbital expansion that strictly follows the momentum conservation equation $\dot{a}/a = - \dot{M}_{\mathrm{tot}}/M_{\mathrm{tot}}$, where $a$ and $\dot{a}$ denote the orbital semi-major axis and its time derivative, respectively, while $M_{\mathrm{tot}}$ is the total mass of the evolving system and $\dot{M}_{\mathrm{tot}}$ represents the system's total mass loss rate. This continuous mass loss alters the semi-major axis ratio between the outer and inner orbits, profoundly impacting the system's long-term gravitational stability. \texttt{TSE} monitors the dynamical state in real-time using the empirical stability criterion proposed by \cite{2001MNRAS.321..398M}. A dynamically stable system must satisfy the following proportionality relationship between the outer orbit's periapsis distance and the inner orbit's semi-major axis:
\begin{equation}
	\frac{a_{\mathrm{out}}}{a_{\mathrm{in}}} > 2.8 \left( 1 + \frac{M_3}{M_1+M_2} \right)^{2/5} \frac{(1+e_{\mathrm{out}})^{2/5}}{(1-e_{\mathrm{out}})^{6/5}} \left( 1 - \frac{0.3 i}{180^{\circ}} \right),
\end{equation}
where $a_{\mathrm{out}}$ and $a_{\mathrm{in}}$ denote the semi-major axes of the outer and inner orbits, respectively; $M_1$ and $M_2$ represent the masses of the two inner binary components; $M_3$ is the mass of the outer tertiary companion; $e_{\mathrm{out}}$ is the eccentricity of the outer orbit; and $i$ is the mutual inclination angle between the inner and outer orbital planes, expressed in degrees.

\subsection{Chemically Homogeneous Evolution}

CHE represents one of the crucial physical channels for explaining the formation of massive BBH mergers \cite{2016MNRAS.458.2634M,2023ApJ...952...79L,2016A&A...588A..50M}. This type of evolution typically occurs in extremely close and rapidly rotating massive binary systems. Driven by intense rotational mixing, the stellar interiors achieve nearly complete chemical homogeneity, allowing hydrogen to be almost entirely burned into helium during the main sequence phase \cite{2006A&A...460..199Y, 2009A&A...497..243D, 2023MNRAS.526.4335W}. 

Unlike standard stellar evolution models, stars undergoing CHE lack an expanding hydrogen-rich envelope. Consequently, their physical radii remain largely constant throughout the main sequence. This characteristic allows them to effectively avoid premature Roche-lobe overflow and the complex mass-exchange processes associated with common-envelope (CE) phases \cite{2016A&A...588A..50M, 2016MNRAS.458.2634M, 2026A&A...706A.105L,2024ApJ...975L...8L,2025ApJ...979L..37L, 2016MNRAS.460.3545D, 2021MNRAS.505..663R, 2023RAA....23b5021Z}.

In this study, we closely adopt the physical constraints and evolutionary parameters for CHE utilized by Li et al. \cite{2025PhRvD.112j3005L} in their investigation of BBH mass distributions. Building upon this, we innovatively incorporate the complete CHE evaluation module and parameter scheme into the TSE code \cite{2022MNRAS.516.1406S}. 

\subsection{Triple Common Envelope}\label{subsec:bps}

In a hierarchical triple system, when the outer tertiary star (Star 3) expands during its stellar evolution, it may eventually fill its Roche lobe. The effective Roche lobe radius of the outer tertiary, $R_{L,3}$, relative to the inner binary is analytically approximated using the classical formula by \cite{1983ApJ...268..368E}:
\begin{equation}
	\label{eq:RL3}
	R_{\mathrm{L},3} = a_{\mathrm{out}} f(q_{\mathrm{out}}), \quad \text{with} \quad f(q) = \frac{0.49 q^{2/3}}{0.6 q^{2/3} + \ln(1 + q^{1/3})},
\end{equation}
where $f(q)$ is the dimensionless effective Roche lobe radius function, and $q_{\mathrm{out}} = M_3 / (M_1 + M_2)$ is the mass ratio of the tertiary to the inner binary. When the tertiary radius exceeds $R_{L,3}$, the stability of the ensuing mass transfer is governed by a critical mass ratio, $q_c$. Following the criterion proposed by \cite{2002MNRAS.329..897H}, dynamically unstable mass transfer occurs if $q_{\mathrm{out}} > q_c$. For giant donors with well-developed convective envelopes, $q_c$ is analytically approximated as:
\begin{equation}
	\label{eq:qc}
	q_c = 0.362 + \left[ 3 \left( 1 - \frac{M_{c,3}}{M_3} \right) \right]^{-1},
\end{equation}
where $M_{c,3}$ and $M_3$ are the core mass and total mass of the tertiary star, respectively. 

If this dynamical instability criterion is met, the ensuing mass transfer triggers a TCE phase, where the tertiary's envelope rapidly expands and completely engulfs the inner binary \cite{2013A&ARv..21...59I, 2016ComAC...3....6T}. Because traditional stellar evolution codes currently lack a self-consistent physical module for TCE evolution, we compute the subsequent orbital decay using the recently developed Single Components' Angular momenTum TransfER (\texttt{SCATTER}) CE formalism specifically designed for triple systems \cite{2026MNRAS.547ag192D}. Unlike traditional energy-conservation models, the \texttt{SCATTER} mechanism is grounded in the conservation of total angular momentum. By introducing a mass-ratio-dependent angular momentum transfer function, this mechanism allows each component in the stellar system to independently exchange angular momentum with the ejected envelope. According to this theoretical framework, the post-CE shrinkage of the outer orbit is given by \cite{2023ApJ...944...87D}:
\begin{equation}
	\begin{split}
		\frac{a_{\mathrm{out}}(f)}{a_{\mathrm{out}}(0)} &= \left(\frac{M_{\mathrm{binary}} + M_3^c}{M_{\mathrm{binary}} + M_3(0)}\right) \left(\frac{M_3(0)}{M_3^c}\right)^2 \\
		&\quad \times \exp\left[-\frac{2 M_3^{\mathrm{env}}}{M_{\mathrm{binary}} + M_3^c}\mathcal{F}(M_3^c/M_{\mathrm{binary}})\right],
	\end{split}
\end{equation}
where $M_{\mathrm{binary}}$ is the total mass of the inner binary (assumed constant during the CE process), $M_3(0)$ is the initial mass of Star 3, and $M_3^c$ is its core mass. The function $\mathcal{F}$ characterizes the efficiency of angular momentum transfer and is defined as:
\begin{equation}
	\mathcal{F}(q_{c,j}) = \frac{\eta}{q_{c,j}}\mathcal{Q}(q_{c,j}) + \frac{\eta}{q_{j,c}}\mathcal{Q}(q_{j,c}),
\end{equation}
Here, $\eta$ is the angular momentum transfer efficiency parameter. Through fitting a sample of known post-CE binaries, it is found to satisfy the relationship:
\begin{equation}
	\log_{10}[\eta] = -A \log_{10}\left[\frac{M^{\mathrm{interact}}}{M_{\mathrm{tot}}(f)}\right] + B,
\end{equation}
Typical values are taken as $A=0.95$ and $B=0.6$ \cite{2023ApJ...944...87D}. The function $\mathcal{Q}$ describes the distribution of the envelope mass between the two inspiraling stars, based on the standard Roche lobe approximation \cite{1983ApJ...268..368E}:
\begin{equation}
	\mathcal{Q}(q_{c,j}) = \frac{f(q_{c,j})^\delta}{f(q_{c,j})^\delta + f(q_{j,c})^\delta},
\end{equation}
where $f(q)$ is the dimensionless Roche lobe radius function defined earlier. Typically, $\delta=3$ is adopted because it corresponds to an orbital volume scaling and reduces the sensitive dependence on the value of $\eta$. 

Meanwhile, the orbit of the inner binary also shrinks due to interaction with a portion of the envelope material. The effective envelope mass that exchanges angular momentum with the inner binary is $M_{\mathrm{in}}^{\mathrm{interact}}$, given by:
\begin{equation}
	M_{\mathrm{in}}^{\mathrm{interact}} = M_3^{\mathrm{env}} \times \mathcal{Q}(M_{\mathrm{binary}}/M_3^c),
\end{equation}
The change in the inner orbit is:
\begin{equation}
	\frac{a_{\mathrm{in}}(f)}{a_{\mathrm{in}}(0)} = \exp\left[-\frac{2 M_{\mathrm{in}}^{\mathrm{interact}}}{M_{\mathrm{binary}}}\mathcal{F}(M_1/M_2)\right],
\end{equation}
Accurately modeling these orbital changes is crucial for understanding the formation channels of merging compact binaries and subsequent gravitational wave events \cite{2016Natur.534..512B, 2016PhRvL.116f1102A, 2017ApJ...841...77A}.

\subsection{Natal Kick }

At the terminal stages of massive star evolution, the magnitude of the natal kick serves as a crucial parameter determining the survival of compact binary systems and their subsequent gravitational-wave merger rates. This kick velocity exhibits a significant dependence on the progenitor mass. According to extensive theoretical and population synthesis studies \cite{2001ApJ...554..548F, 2015MNRAS.451.4086S}, the physical magnitude of a black hole's natal kick is primarily governed by the depth of matter fallback during the supernova explosion.  Mathematically, the effective natal kick velocity $v_{\mathrm{nk}}$ imparted to a newly formed black hole is commonly modeled by modulating a standard neutron star kick distribution by the fallback fraction $f_{\mathrm{fb}}$ \cite{2012ApJ...749...91F, 2013ApJ...779...72D}:
\begin{equation}
	v_{\mathrm{nk}} = (1 - f_{\mathrm{fb}}) v_{\mathrm{max}},
\end{equation}
where $v_{\mathrm{max}}$ is the velocity drawn from a Maxwellian distribution typical of neutron stars \cite{2005MNRAS.360..974H}, and $f_{\mathrm{fb}} \in [0,1]$ represents the fraction of the stellar envelope that falls back onto the proto-compact object. Low-mass progenitors trigger energetic explosions with minimal fallback ($f_{\mathrm{fb}} \to 0$), thereby imparting the compact object with a high-velocity recoil comparable to the observed proper motions of pulsars \cite{2012ARNPS..62..407J}. Conversely, as the progenitor mass increases, the enhanced gravitational binding energy leads to massive, symmetric fallback ($f_{\mathrm{fb}} \to 1$) that effectively cancels out the imparted momentum. Furthermore, classical studies on the fate of massive stars \cite{2003ApJ...591..288H}, later adopted by dynamical models \cite{2017ApJ...841...77A}, indicate that in a solar-metallicity environment, stars with a zero-age main sequence (ZAMS) mass of $m \gtrsim 40 M_{\odot}$ undergo a direct collapse scenario ($f_{\mathrm{fb}} = 1$), where the massive fallback process is sufficient to completely suppress any natal kick.

\subsection{Black Hole Spin and Gravitational Recoil }

The dimensionless spin vector of a black hole, $\vec{\chi} = c\vec{S}/(GM^2)$ (where $\vec{S}$ is the spin angular momentum), plays a pivotal role in governing the dynamical evolution and gravitational-wave emission of compact binaries. According to recent stellar evolution models incorporating efficient angular momentum transport mechanisms---such as the Tayler-Spruit dynamo \cite{2002A&A...381..923S, 2019ApJ...881L...1F}---the cores of massive stars are expected to lose the vast majority of their angular momentum prior to collapse. Consequently, first-generation (1g) black holes originating from isolated stellar evolution are generally assumed to possess negligible natal spins \cite{2020A&A...636A.104B}. In our framework, we adopt this standard paradigm, setting the initial dimensionless spin magnitudes strictly to zero ($\vec{\chi}_1 = \vec{\chi}_2 = 0$).

Upon the coalescence of a BBH system, the resulting remnant black hole acquires a final dimensionless spin, $\chi_{\mathrm{rem}}$, which is dictated by the orbital angular momentum at the innermost stable circular orbit (ISCO) alongside the initial spins of the progenitors. For initially non-spinning BBHs, the spin of the merger remnant is generated entirely by the orbital angular momentum. Based on extensive numerical relativity simulations and their associated fitting formulas \cite{2008ApJ...679.1422R, 2009ApJ...704L..40B, 2016ApJ...825L..19H, 2017PhRvD..95f4024J}, the magnitude of the remnant spin can be analytically expressed as a polynomial function of the symmetric mass ratio $\eta = q/(1+q)^2$ (where $q = m_2/m_1 \leq 1$):
\begin{equation}
	\chi_{\mathrm{rem}} = 2\sqrt{3}\eta - 0.434\eta^2 - 3.280\eta^3 + 3.328\eta^4,
\end{equation}
Under this formulation, the coalescence of an equal-mass, non-spinning binary ($q=1$, $\eta=0.25$) yields a highly spinning second-generation (2g) black hole with a dimensionless spin of $\chi_{\mathrm{rem}} \approx 0.686$ \cite{2006PhRvD..73f1501C}.

During the inspiral, merger, and ringdown phases of a BBH coalescence, the anisotropic emission of gravitational radiation carries away a net linear momentum. To conserve momentum, the merger remnant receives an instantaneous impulse in the opposite direction---a phenomenon known as gravitational recoil or kick \cite{2006ApJ...653L..93B, 2016PhRvD..93l4066G}. Based on numerical relativity simulations, the recoil velocity vector $\vec{v}_{\mathrm{kick}}$ is generally parameterized as the sum of mass-asymmetry and spin-asymmetry contributions \cite{2007PhRvL..98i1101G, 2007PhRvL..98w1102C, 2012PhRvD..85h4015L}:
\begin{equation}
	\vec{v}_{\mathrm{kick}} = v_m \hat{e}_{\perp, 1} + v_{\perp} (\cos\xi \hat{e}_{\perp, 1} + \sin\xi \hat{e}_{\perp, 2}) + v_{\parallel} \hat{e}_{\parallel},
\end{equation}
where $v_m \propto \eta^2 \sqrt{1-4\eta}$ represents the mass-ratio-driven kick within the orbital plane, while $v_{\perp}$ and $v_{\parallel}$ arise from the progenitor spin components parallel and in-plane to the orbital angular momentum, respectively. As established above, we assume our 1g progenitor black holes are born non-spinning ($\vec{\chi}_1 = \vec{\chi}_2 = 0$), causing the spin-asymmetry terms ($v_{\perp}$ and $v_{\parallel}$) to vanish identically. Consequently, the gravitational recoil for these 1g mergers is entirely governed by the mass-asymmetry term $v_m$. Crucially, for our highly synchronized, equal-mass inner binary ($q=1$), $v_m$ naturally drops to zero. This ensures that the newly formed massive 2g black hole experiences absolutely no natal kick ($\vec{v}_{\mathrm{kick}} \approx 0$), perfectly preserving the dynamical integrity of the remaining binary system and facilitating the subsequent interaction with the tertiary companion.


\subsection{Cosmic Integration and Detection Rates}\label{subsec:cosmic}

To quantitatively evaluate the theoretical occurrence and detection rates of BBH mergers, we must consider two primary factors. Firstly, the distribution of delay times introduces significant complexity. While massive binaries typically evolve into BBHs within ~10 Myr, the subsequent merger delay times can span many Gyr \cite{1964PhRv..136.1224P, 2017MNRAS.472.2422M, 2019MNRAS.482..870E}. This makes it difficult to accurately determine the exact redshift $z$ at which the BBH systems originally formed. Because the cosmic star formation rate density (SFRD) varies strongly with redshift \cite{2014ARA&A..52..415M}, this delay-time uncertainty contributes to the imprecision in predicting overall BBH merger rates \cite{2022ApJ...931...17V}.

Secondly, the BBH formation rate is highly metallicity-dependent, with a significantly higher yield in low-metallicity environments \cite{2018MNRAS.474.2959G, 2018MNRAS.480.2011G, 2022MNRAS.516.5737B}. Furthermore, metallicity dictates the stellar mass loss rates, which critically impacts the final remnant masses and the resulting merger rate \cite{2012ARA&A..50..107L, 2024ApJ...976...23B}. 

To account for these effects, our study adopts the metallicity-specific SFRD, $\mathrm{SFRD}(Z_{\rm i}, z)$, from \cite{2019MNRAS.490.3740N}. They decouple the calculation of the SFRD into two independent factors: the overall SFR and the metallicity density function (${\rm d}P/{\rm d}Z$), such that:
\begin{equation}
\frac{\mathrm{d}^3 M_{\mathrm{SFR}}}{\mathrm{~d} t_s \mathrm{~d} V_c \mathrm{~d} Z}(z)=\frac{\mathrm{d}^2 M_{\mathrm{SFR}}}{\mathrm{~d} t_s \mathrm{~d} V_c}(z) \times \frac{\mathrm{d} P}{\mathrm{~d} Z}(z),
\end{equation}
In this framework, the total SFRD follows the parametric form proposed by \cite{2014ARA&A..52..415M} and \cite{2017ApJ...840...39M}:
\begin{equation}
	\psi(z) \equiv \frac{\mathrm{d}^2 M_{\mathrm{SFR}}}{\mathrm{d} t_s \mathrm{d} V_c} = a \frac{(1+z)^b}{1+[(1+z) / c]^d} \, \mathrm{M}_{\odot} \, \mathrm{yr}^{-1} \, \mathrm{Mpc}^{-3},
\end{equation}
where the fitting parameters are $a = 0.015$, $b = 2.7$, $c = 2.9$, and $d = 5.6$. For the metallicity density function (${\rm d}P/{\rm d}Z$).

By applying this $\mathrm{SFRD}(Z_{\rm i}, z)$ with sampling weights adjusted according to the assumed cosmic metallicity distribution, the intrinsic BBH merger rate density, $\mathcal{R}_{\mathrm{m}}$, is obtained by integrating the BBH formation rate density over all metallicities and delay times:
\begin{equation}
\begin{aligned}
	& \mathcal{R}_{\mathrm{m}}\left(t_{\mathrm{m}}, M_{\mathrm{l}}, M_2\right) \equiv \frac{d^4 N_{\text {merger }}}{d t_{\mathrm{m}} d V_{\mathrm{c}} d M_1 d M_2}\left(t_{\mathrm{m}}, M_{\mathrm{l}}, M_2\right) \\
	& =\int d Z_{\mathrm{i}} \int_0^{t_{\mathrm{m}}} d t_{\text {delay }} \operatorname{SFRD}\left(Z_{\mathrm{i}}, z\left(t_{\text {form }}=t_{\mathrm{m}}-t_{\text {delay }}\right)\right) \\
	& \times \frac{d^4 N_{\text {form }}}{d M_{\mathrm{SFR}} d t_{\text {delay }} d M_1 d M_2}\left(Z_{\mathrm{i}}, t_{\text {delay }}, M_{\mathrm{l}}, M_2\right),
\end{aligned}
\end{equation}
where $V_{\mathrm{c}}$ is the comoving volume, and the formation time $t_{\rm form}$ is related to the merger time $t_{\rm m}$ and delay time $t_{\rm delay}$ by $t_{\rm form}=t_{\rm m}-t_{\rm delay}$. The final term in this equation represents the formation rate of merging BBHs per unit of stellar mass formed, evaluated at a given initial metallicity $Z_{\mathrm{i}}$.

Following \cite{2019MNRAS.490.3740N}, the local differential detection rate $\mathcal{R}_{\mathrm{det}}$ for the Cosmic Integration pipeline is given by:
\begin{equation}
\begin{aligned}
	& \mathcal{R}_{\mathrm{det}}\left(t_{\mathrm{det}}, M_1, M_2\right) \equiv \frac{d^3 N_{\mathrm{det}}}{d t_{\mathrm{det}} d M_1 d M_2} \\
	& =\int d z \frac{d V_{\mathrm{c}}}{d z} \frac{d t_{\mathrm{m}}}{d t_{\mathrm{det}}} \mathcal{R}_{\mathrm{m}}\left(t_{\mathrm{m}}\right) P_{\mathrm{det}}\left(M_1, M_2, z\left(t_{\mathrm{m}}\right)\right),
\end{aligned}
\end{equation}
where $t_{\mathrm{det}}$ is the time in the observer (detector) frame, $z$ is the redshift, $V_{\mathrm{c}}$ is the comoving volume, and $\mathcal{R}_{\mathrm{m}}$ is the intrinsic merger rate derived above. $P_{\mathrm{det}}$ is the probability of detecting a gravitational-wave signal from a binary with component masses $M_{\rm 1}$ and $M_{\rm 2}$ merging at redshift $z$.

Theoretically, the expected number of detected events ($N_{\mathrm{obs}}$) for a specific sub-population is dictated by the intrinsic volumetric merger rate density ($\mathcal{R}_{\mathrm{m}}$) integrated over the accessible cosmological volume. Incorporating the cosmological time dilation factor $(1+z)^{-1}$ \cite{2023PhRvX..13a1048A, 2018ApJ...863L..41F}, the effective sensitive spacetime volume ($\langle V T \rangle$) for a sub-population with intrinsic parameters $\theta$ is defined as:
\begin{equation}
	\langle V T \rangle = T_{\mathrm{obs}} \int_{z_{\min}}^{z_{\max}} \frac{1}{1+z} \left( \frac{dV_c}{dz} \right) p_{\mathrm{det}}(z, \theta) \, dz,
\end{equation}
where $p_{\mathrm{det}}(z, \theta)$ encapsulates the network detection probability, accounting for instrumental selection effects such as signal-to-noise ratio (SNR) thresholds and antenna pattern variations.

\section{Results}\label{sec:results}
We filter the gravitational-wave catalog to identify events satisfying the following stringent criteria: a probability of astrophysical origin ($p_{\mathrm{astro}}$) greater than 0.9, a primary black hole mass residing within the PISN mass gap ($m_1 > 45 M_{\odot}$), a secondary black hole mass in the standard stellar-mass regime ($m_2 < 45 M_{\odot}$), and the presence of a high effective spin ($\chi_{\mathrm{eff}} > 0.2$). The initial sample satisfying these parameter constraints includes GW230107, GW230928, GW230820, GW190706, GW190620, and GW170729. After meticulously excluding events that exhibit negative effective spins, we ultimately find that the remaining sources, notably GW230824 and GW190706, can be exceptionally well-explained by our proposed hierarchical triple evolution model.

\begin{figure*}[htbp]
	\centering
	\includegraphics[width=\textwidth, height=0.85\textheight, keepaspectratio]{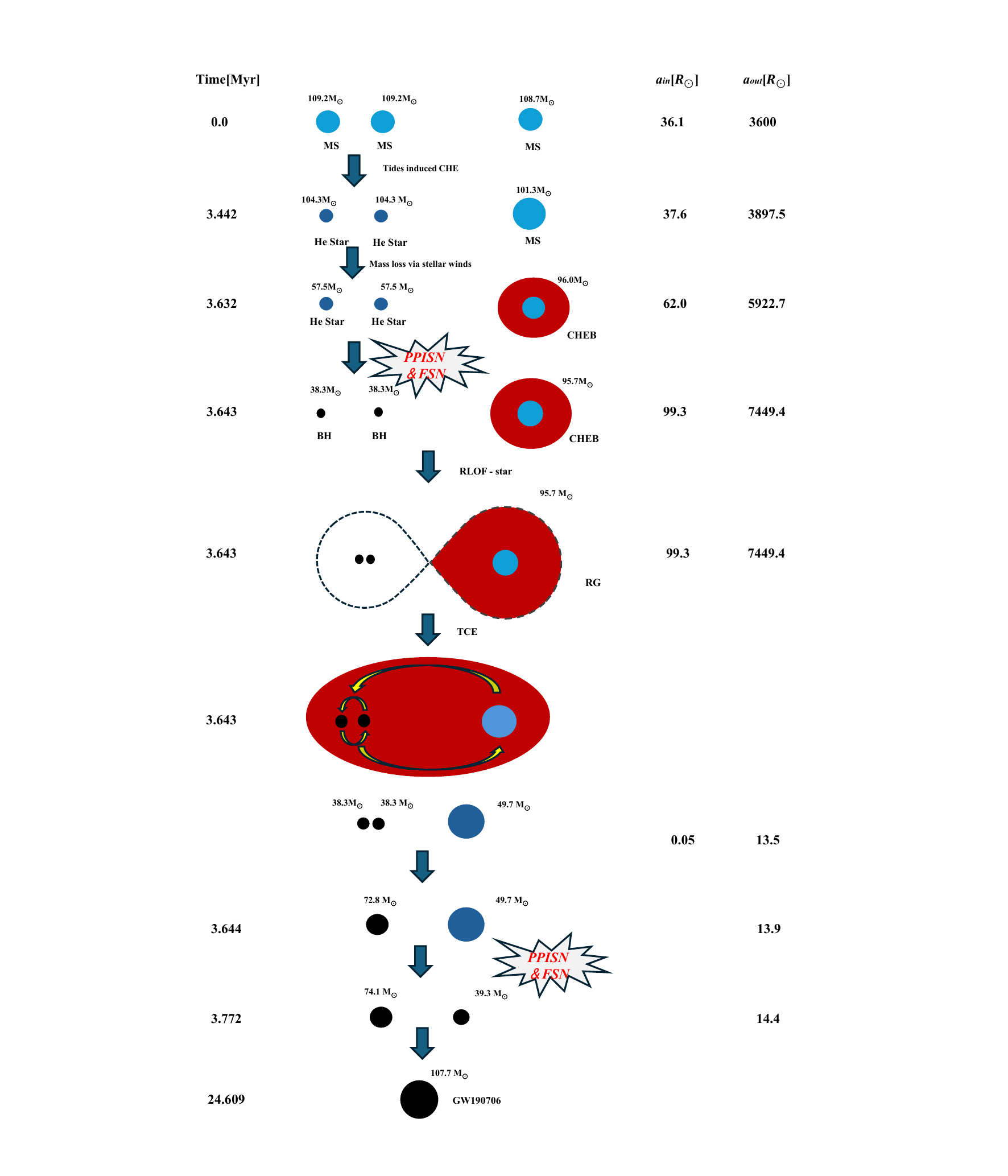}
	\caption{The complete evolutionary pathway of GW190706, from the zero-age main sequence (ZAMS) to the final merger.}
	\label{fig:GW190706}
\end{figure*}

 \subsection{A Formation Pathway of GW190706}

GW190706 is a remarkable BBH merger event cataloged in GWTC-3, featuring component masses of approximately $74.0_{-16.9}^{+20.1} \, M_{\odot}$ and $39.4_{-15.4}^{+18.4} \, M_{\odot}$ \cite{2023PhRvX..13d1039A}. The primary black hole explicitly resides within the PISN mass gap, posing a significant challenge to standard isolated binary evolution models. Here, we demonstrate that the hierarchical triple evolution framework provides a highly consistent and natural explanation for the origin of GW190706.

Figure \ref{fig:GW190706} illustrates the kinematic diagram of key evolutionary stages, as well as the temporal evolution of the orbital radii, stellar masses, and stellar radii. In our fiducial configuration, the system is initialized as a coplanar hierarchical triple situated in a low-metallicity environment ($Z = 0.001$). While massive binaries in such extremely tight orbits may initially form with unequal masses ($q_{\mathrm{in}} \neq 1$), early mass transfer during an overcontact phase is expected to rapidly drive the mass ratio toward unity. Therefore, to facilitate the modeling of the subsequent evolutionary phases, we assume this equilibration has already occurred and initialize the inner binary with a mass ratio of exactly unity ($M_1 = M_2 = 109.2 \, M_{\odot}$). The distant outer tertiary has a mass of $M_3 = 108.7 \, M_{\odot}$. The initial inner and outer orbital separations are $a_{\mathrm{in}} = 36.1 \, R_{\odot}$ and $a_{\mathrm{out}} = 3600 \, R_{\odot}$, respectively, both initialized with zero eccentricity ($e_{\mathrm{in}} = e_{\mathrm{out}} = 0$).

The extreme tightness of the inner orbit dictates its subsequent evolution via strong tidal interactions. While the massive stars initially possess typical baseline rotational velocities at the zero-age main sequence, intense tidal torques rapidly spin up the inner binary components. As tidal synchronization enforces orbital and spin equilibrium, the spin angular velocities of the stars ($\omega_{\mathrm{spin}} \approx 1352 \, \mathrm{rad/yr}$) drastically exceed the mass- and metallicity-dependent critical threshold required for rotational mixing ($\omega_{\mathrm{crit}} \approx 1119 \, \mathrm{rad/yr}$, yielding a ratio of $\omega_{\mathrm{spin}} / \omega_{\mathrm{crit}} \approx 1.21$). Surpassing this critical bifurcation point triggers highly efficient internal meridional circulation. Consequently, they bypass the traditional Hertzsprung-Russell evolutionary track and instead undergo CHE. 

These stars evolve directly from the main sequence to naked helium stars, entirely bypassing the traditional giant phases. During this helium star phase, intense stellar-wind mass loss substantially reduces the masses of the inner binary components \cite{2023A&A...674A.216L}, leading to an adiabatic widening of both the inner and outer orbits. For those with pre-supernova masses falling into the aforementioned PPISN regime (typically $\sim 32\text{--}64 \, M_{\odot}$), subsequent violent pulsations strip away a significant fraction of their outer layers. As demonstrated in our evolutionary models, a naked helium star reaching a pre-supernova mass of approximately $57.5 \, M_{\odot}$ ejects roughly $19.2 \, M_{\odot}$ of material during this phase. Once these mass loss episodes cease, the remnant core remains too massive to drive a successful supernova shockwave. Ultimately, they collapse directly into black holes via failed supernovae(FSNe) \cite{2023ApJ...952...79L, 2025A&A...704A..46L}.

Following the formation of the inner BBH, the outer tertiary star continues its stellar evolution. As it evolves into a giant phase, it expands and eventually fills its Roche lobe ($R_{\mathrm{L},3}$, defined in Equation (\ref{eq:RL3})). When the tertiary radius exceeds $R_{\mathrm{L},3}$, the stability of the ensuing mass transfer is governed by a critical mass ratio, $q_c$. According to \cite{2002MNRAS.329..897H}, dynamically unstable mass transfer occurs if the outer mass ratio exceeds this threshold ($q_{\mathrm{out}} > q_c$). For giant donors with well-developed convective envelopes, $q_c$ is analytically approximated as Equation (\ref{eq:qc}). This dynamically unstable mass transfer triggers a TCE phase. We compute the subsequent orbital decay using the \texttt{SCATTER} formalism discussed in Section \ref{subsec:bps}.

Upon the successful ejection of the envelope, we employ the revised gravitational-wave radiation timescale formula recently proposed by \cite{2020MNRAS.495.2321Z}, which incorporates first-order post-Newtonian (1PN) corrections and self-consistent eccentricity evolution into the classical \cite{1964PhRv..136.1224P} estimate. The orbital decay timescale of the post-CE inner BBH is evaluated as:
\begin{equation}
	t_{\mathrm{GW}} = \frac{5}{256} \frac{c^5}{G^3} \frac{a_{\mathrm{in}, f}^4}{M_1 M_2 (M_1 + M_2)} \times R(e_{\mathrm{in}, f}) \times Q_f(p_{\mathrm{in}, f}),
\end{equation}
where $a_{\mathrm{in}, f}$, $e_{\mathrm{in}, f}$, and $p_{\mathrm{in}, f} = a_{\mathrm{in}, f}(1-e_{\mathrm{in}, f})$ are the final semi-major axis, eccentricity, and periapsis of the inner orbit post-TCE, respectively. The correction factors $R$ and $Q_f$ are analytically defined as:
\begin{equation}
	R(e_{\mathrm{in}, f}) = 8^{1 - \sqrt{1 - e_{\mathrm{in}, f}}}, \quad Q_f(p_{\mathrm{in}, f}) = \exp\left( 2.5 \frac{r_S}{p_{\mathrm{in}, f}} \right),
\end{equation}
with $r_S = 2G(M_1 + M_2)/c^2$ representing the Schwarzschild radius of the inner binary. Even though our inner binary remains nearly circular after the extensive tidal synchronization and CE phases ($e_{\mathrm{in}, f} \approx 0 \implies R \approx 1$), the 1PN correction factor $Q_f$ provides a significantly more robust timescale for such extreme massive BBHs approaching the innermost stable circular orbit. If the evaluated timescale yields $t_{\mathrm{GW}} \le 1 \, \mathrm{yr}$, we determine that a prompt merger of the inner binary occurs, effectively reducing the triple system to a new binary. 

Following the merger, the mass of the newly formed black hole is determined using contemporary numerical relativity results, assuming a radiative loss of approximately $5\%$ of the system's total mass \cite{2010PhRvD..82f4016S}. 

Crucially, because the inner BBH has a mass ratio of $q=1$ and their spins are perfectly aligned (a direct consequence of prior tidal synchronization and CHE), the merger produces absolutely zero gravitational recoil kick ($\vec{v}_{\mathrm{kick}} \approx 0$). This guarantees that the newly formed massive 2g black hole is not ejected, securely maintaining the dynamical integrity of the remaining binary system. 

At this juncture, the surviving system consists of this massive 2g remnant---securely located within the PISN mass gap---paired with the stripped helium core of the tertiary star. We meticulously record the epoch when the tertiary helium core emerges and the termination time of the TCE to estimate its remaining evolutionary timescale. Similar to the inner components, this tertiary helium core also experiences violent episodic mass loss via PPISNe before undergoing a FSNe, ultimately collapsing directly to form a secondary black hole. Because all black holes in this system form via direct collapse without strong natal kicks, the orbit preserves a highly coplanar configuration. 

Assuming zero-spin 1g BHs, the massive 2g remnant inherits a substantial dimensionless spin ($a \approx 0.686$) purely from the inner orbital angular momentum. When paired with the zero-spin tertiary 1g black hole, this strictly coplanar topology keeps the spins perfectly aligned, naturally producing a large positive effective inspiral spin ($\chi_{\text{eff}} \approx 0.45$). Driven by subsequent gravitational-wave radiation, the orbit of this newly formed, highly asymmetric BBH system decays until coalescence, remarkably reproducing the observable properties of the GW190706 event. Specifically, gravitational-wave observations constrain the primary and secondary masses of GW190706 to $m_1 = 74.0_{-16.9}^{+20.1} \, M_{\odot}$ and $m_2 = 39.4_{-15.4}^{+18.4} \, M_{\odot}$ with an effective inspiral spin of $\chi_{\text{eff}} = 0.28_{-0.31}^{+0.25}$ \cite{2023PhRvX..13d1039A}, which are perfectly consistent with our dynamically assembled mass predictions of roughly $72.8 \, M_{\odot}$ and $39.3 \, M_{\odot}$ alongside our predicted effective spin. This striking consistency demonstrates that such hierarchical mergers of stripped helium stars provide a natural and compelling astrophysical channel for massive, highly spinning black hole binaries located within the PISN mass gap.

\subsection{Initial Parameter Distributions and System Probability}

To strictly constrain the initial parameter space for our hierarchical triple channel, we adopt distributions motivated by both theoretical requirements for PISN mass gap black holes and recent empirical observations of massive multiple-star systems. To completely avoid severe stellar-wind mass stripping prior to core collapse, the ambient metallicity is fixed at an extremely metal-poor value of $Z = 0.001$.

The primary mass $M_1$ is sampled first from the standard initial mass function (IMF) \cite{2001MNRAS.322..231K} with the probability density function $P(M_1) \propto M_1^{-2.3}$ for $M_1 > 0.1 M_{\odot}$. Normalizing the IMF over a standard stellar mass range of $M_1 \in [0.1, 150] \, M_{\odot}$, we tightly constrain the sampling interval to $\Delta M_1 \in [107, 140] \, M_{\odot}$ to ensure the remnant successfully enters the PISN mass gap. Integrating the IMF over this specific mass window yields a probability component of $P_{M_1} \approx 7.47 \times 10^{-5}$. Furthermore, given that extreme massive O-type stars almost exclusively reside in multiple systems, we incorporate a realistic hierarchical triple fraction of $f_{\mathrm{trip}} = 0.73$ based on comprehensive observational statistics (\cite{ 2017ApJS..230...15M}).

For the inner binary, we follow the covariant sampling framework of \cite{2017ApJS..230...15M}, where the orbital period must be established prior to the mass ratio, as the latter strongly depends on the former. The initial inner orbital period is restricted to an extremely tight configuration of $\Delta P_{\mathrm{in}} \in [1.2, 2.0]$ days ($\Delta \log_{10} P \approx 0.222$ dex), facilitating rapid tidal synchronization and CHE. Adopting the empirically observed period distribution density for close massive binaries ($f(\log_{10} P) \approx 0.24 \, \mathrm{dex}^{-1}$), this yields a probability of $P_{P_{\mathrm{in}}} \approx 0.053$. Conditioned strictly on this short orbital period, the inner mass ratio is subject to a pronounced ``twin excess''. Consequently, the conditional probability of drawing an inner mass ratio within the peak interval $\Delta q_{\mathrm{in}} \in [0.9, 1.0]$ is estimated to be $P_{q_{\mathrm{in}}} \approx 0.150$.

Similarly, for the outer tertiary star, the orbital period $P_{\mathrm{out}}$ is determined before sampling its mass. Over the required wide interval of $\Delta P_{\mathrm{out}} \in [400, 1200]$ days ($\Delta \log_{10} P \approx 0.477$ dex), assuming a period probability density of roughly $0.1 \, \mathrm{dex}^{-1}$ for wide massive companions, we obtain $P_{P_{\mathrm{out}}} \approx 0.048$. Given this wide orbital configuration, \cite{2017ApJS..230...15M} demonstrate that the mass ratio distribution shifts away from the twin excess and instead follows a power-law $f(q_{\mathrm{out}}) \propto q_{\mathrm{out}}^{-1.7}$. Normalizing this distribution over $q_{\mathrm{out}} \in [0.1, 1.0]$, the conditional probability for sampling a tertiary mass ratio within $\Delta q_{\mathrm{out}} \in [0.4, 0.5]$ is $P_{q_{\mathrm{out}}} \approx 0.068$.

Based on the covariant integration framework established above, the comprehensive birth probability of triple systems fulfilling these rigorous initial conditions is the direct product of these sequentially dependent components:
\begin{equation}
	\begin{split}
		P_{\mathrm{sys}} &= f_{\mathrm{trip}} \cdot P_{M_1} \cdot P_{P_{\mathrm{in}}} \cdot P_{q_{\mathrm{in}}} \cdot P_{P_{\mathrm{out}}} \cdot P_{q_{\mathrm{out}}} \\
		&\approx 1.41 \times 10^{-9},
	\end{split}
\end{equation}
This rigorously derived probability ($P_{\mathrm{sys}}$) serves as the fundamental basis for our subsequent volumetric merger rate calculations.

\subsection{Intrinsic Merger Rate Density and Observational Consistency}\label{subsec:merger_rate}

Having rigorously integrated the initial parameter distributions and accounting for the multiplicity fraction of massive stars ($f_{\mathrm{trip}} \approx 0.73$, \cite{2017ApJS..230...15M}), we obtain the comprehensive birth probability for our hierarchical triple systems to be $P_{\mathrm{sys}} \approx 1.41 \times 10^{-9}$.

To quantify the efficiency of this channel, we calculate the intrinsic merger rate density, $\mathcal{R}_{\mathrm{triple}}$. Given that the timescale from formation to merger for these systems is typically on the order of 10 Myr, we neglect the delay-time distribution and approximate the rate as being instantaneous with respect to the star formation rate. Adopting a fiducial survival fraction of $f_{\mathrm{surv}} = 0.5$ to account for dynamical disruption, and evaluating the cosmic star formation rate $\psi(z \approx 0.68) \approx 0.058 \, \mathrm{M_{\odot} \, yr^{-1} \, Mpc^{-3}}$ alongside a metallicity weight factor $\frac{\mathrm{d} P}{\mathrm{d} Z} \approx 0.16$ (for $Z=0.001$), the merger rate density is expressed as:
\begin{equation}
	\begin{split}
		\mathcal{R}_{\mathrm{triple}}(z=0.68) &= \frac{\psi(0.68)}{\langle M \rangle} \cdot \frac{\mathrm{d} P}{\mathrm{~d} Z}(z) P_{\mathrm{sys}} f_{\mathrm{surv}} \times 10^9 \\
		&\approx 0.011 \, \mathrm{Gpc^{-3} \, yr^{-1}},
	\end{split}
\end{equation}

To appropriately contextualize this theoretical prediction, we establish a strictly constrained observational baseline. To ensure a conservative and robust empirical evaluation, we calculate the network detection probability at the most challenging limit of our parameter space ($M_1=50\,M_\odot, M_2=30\,M_\odot, z=0.87$). Following the projection framework of \cite{2015ApJ...806..263D} with a threshold SNR of $\rho_{\mathrm{thr}} = 12$ for a three-detector network, we obtain a minimum average detection probability of $p_{\mathrm{det}} \approx 0.347$. Over the relevant redshift horizon ($z \in [0.49, 0.87]$) and an observation period of $T_{\mathrm{obs}} \approx 7$ years, the time-corrected geometric comoving volume is approximately $345 \, \mathrm{Gpc^3 \, yr}$, resulting in an effective sensitive spacetime volume of $\langle V T \rangle \approx 119.7 \, \mathrm{Gpc^3 \, yr}$. 

By inverting the relationship for our $N_{\mathrm{obs}} = 6$ empirically identified highly-spinning mass gap events, we dynamically derive the empirical volumetric merger rate density:
\begin{equation}
	\mathcal{R}_{\mathrm{obs}} = \frac{N_{\mathrm{obs}}}{\langle V T \rangle} \approx \frac{6}{119.7} \approx 0.050 \, \mathrm{Gpc^{-3} \, yr^{-1}},
\end{equation}

By comparing our theoretically derived rate ($\approx 0.011 \, \mathrm{Gpc^{-3} \, yr^{-1}}$) against this rigorous empirical baseline ($\approx 0.050 \, \mathrm{Gpc^{-3} \, yr^{-1}}$), we find that the isolated hierarchical triple evolution channel can robustly account for approximately $22\%$ of the total observed rate. This substantial theoretical efficiency confirms that massive hierarchical triples operating in metal-poor environments represent a highly significant and fundamental astrophysical origin for these extreme systems. Importantly, a $22\%$ contribution comfortably avoids overpredicting the population, naturally leaving a necessary physical margin for alternative assembly pathways such as dynamical capture in dense star clusters or AGN accretion disks.
\section{Conclusion} \label{sec:conclusion}

In this study, we investigate the astrophysical origins of the extremely massive, highly-spinning BBH mergers located unequivocally within the PISN mass gap. Utilizing the latest LVK gravitational-wave catalogs (GWTC-4), we isolate a specific sub-population of six mass gap events (exemplified by GW190706 and GW230824) characterized by positive effective spins, which strongly points towards a primordial stellar origin rather than isotropic dynamical capture. To explain these extreme phenomena, we propose and quantitatively evaluated an isolated hierarchical triple stellar evolution channel operating in metal-poor environments ($Z=0.001$).

Our evolutionary framework relies on a tightly coupled sequence of physical processes. The extremely tight inner binary ($P_{\mathrm{in}} \in [1.2, 2.0]$ days, $q_{\mathrm{in}} \approx 1$) undergoes rapid tidal synchronization, forcing the massive progenitors into CHE. This process effectively bypasses the traditional giant expansion phase, allowing the inner binary to evolve quiescently into a binary black hole. Crucially, the symmetric mass ratio and aligned spins act in concert to produce a near-zero natal kick, thereby perfectly preserving the integrity of the system. Subsequently, the outer tertiary companion evolves, fills its Roche lobe, and triggers a TCE phase. The successful ejection of this envelope dramatically shrinks the orbits of both the inner BBH and the outer tertiary, rapidly driving the inner binary to coalescence via gravitational-wave radiation.

By rigorously integrating the requisite initial parameter space governed by empirical probability density functions, and adopting a realistic multiple-star fraction of $f_{\mathrm{trip}} = 0.73$, we determine the comprehensive birth probability of these specific triple systems to be $P_{\mathrm{sys}} \approx 1.41 \times 10^{-9}$. Incorporating a survival fraction of $f_{\mathrm{surv}} = 0.5$ against dynamical disruption and a restrictive metallicity weight factor ($\frac{\mathrm{d} P}{\mathrm{~d} Z} \approx 0.16$), we derive a theoretical volumetric merger rate density of $\mathcal{R}_{\mathrm{triple}} \approx 0.011 \, \mathrm{Gpc^{-3} \, yr^{-1}}$ at the representative characteristic redshift ($z \approx 0.68$). 

When contrasted with the empirical observational baseline of $\mathcal{R}_{\mathrm{obs}} \approx 0.050 \, \mathrm{Gpc^{-3} \, yr^{-1}}$ determined by the LVK detector network's sensitivity limitations, we find a highly compelling quantitative alignment. Our isolated hierarchical triple channel can independently robustly account for approximately $22\%$ of the total observed rate for this unique sub-population. 

We conclude that the dynamical perturbation and envelope interactions within massive hierarchical triples serve as a fundamental, highly efficient, and indispensable astrophysical origin for populating the PISN mass gap with highly-spinning black holes. While this channel provides an effective mechanism for generating such extreme systems, it simultaneously preserves a realistic physical margin for alternative pathways, such as hierarchical mergers in dense nuclear star clusters or AGN accretion disks. Future multi-band gravitational-wave observations combining ground-based networks (e.g., Cosmic Explorer, Einstein Telescope) with space-based interferometers (e.g., LISA) will be pivotal in detecting the tertiary companions and definitively confirming the evolutionary demographics of these extraordinary hierarchical systems.

\section*{Acknowledgements}
This work received the support of the National Natural Science Foundation of China under grants 12373038, 12563007, U2031204, and 12288102; the Natural Science Foundation of Xinjiang No. 2022TSYCLJ0006 and 2022001085; the China Manned Space Program grant No. CMS-CSST-2025-A15; and  AP23490322 Exploration of Thermodynamic Properties of Relativistic Compact Objects Within the Framework of Geometrothermodynamics (GTD), Grant financing for scientific and/or scientific-technical projects for 2024-2026 by the Ministry of Science and Higher Education of the Republic of Kazakhstan.

\section*{Data available}

\bibliography{sample631}

@PREAMBLE{ {\providecommand{\noopsort}[1]{}} }

@ARTICLE{2016PhRvL.116f1102A,
	author = {{Abbott}, B.~P. and {Abbott}, R. and {Abbott}, T.~D. and {LIGO Scientific Collaboration} and {Virgo Collaboration}},
	title = "{Observation of Gravitational Waves from a Binary Black Hole Merger}",
	journal = {\prl},
	year = 2016,
	month = feb,
	volume = {116},
	number = {6},
	eid = {061102},
	pages = {061102},
	doi = {10.1103/PhysRevLett.116.061102},
	archivePrefix = {arXiv},
	eprint = {1602.03837},
	primaryClass = {gr-qc},
	adsurl = {https://ui.adsabs.harvard.edu/abs/2016PhRvL.116f1102A}
}

@ARTICLE{2023PhRvX..13d1039A,
	author = {{Abbott}, R. and {Abbott}, T.~D. and {Acernese}, F. and {Zucker}, M.~E. and {Zweizig}, J. and {LIGO Scientific Collaboration} and {VIRGO Collaboration} and {KAGRA Collaboration}},
	title = "{GWTC-3: Compact Binary Coalescences Observed by LIGO and Virgo during the Second Part of the Third Observing Run}",
	journal = {Physical Review X},
	year = 2023,
	month = oct,
	volume = {13},
	number = {4},
	eid = {041039},
	pages = {041039},
	doi = {10.1103/PhysRevX.13.041039},
	archivePrefix = {arXiv},
	eprint = {2111.03606},
	primaryClass = {gr-qc},
	adsurl = {https://ui.adsabs.harvard.edu/abs/2023PhRvX..13d1039A}
}

@ARTICLE{2002ApJ...567..532H,
	author = {{Heger}, A. and {Woosley}, S.~E.},
	title = "{The Nucleosynthetic Signature of Population III}",
	journal = {\apj},
	year = 2002,
	month = mar,
	volume = {567},
	number = {1},
	pages = {532-543},
	doi = {10.1086/338487},
	archivePrefix = {arXiv},
	eprint = {astro-ph/0107037},
	primaryClass = {astro-ph},
	adsurl = {https://ui.adsabs.harvard.edu/abs/2002ApJ...567..532H}
}

@ARTICLE{2019ApJ...887...53F,
	author = {{Farmer}, R. and {Renzo}, M. and {de Mink}, S.~E. and {Marchant}, P. and {Justham}, S.},
	title = "{Mind the Gap: The Location of the Lower Edge of the Pair-instability Supernova Black Hole Mass Gap}",
	journal = {\apj},
	year = 2019,
	month = dec,
	volume = {887},
	number = {1},
	eid = {53},
	pages = {53},
	doi = {10.3847/1538-4357/ab518b},
	archivePrefix = {arXiv},
	eprint = {1910.12874},
	primaryClass = {astro-ph.SR},
	adsurl = {https://ui.adsabs.harvard.edu/abs/2019ApJ...887...53F}
}

@ARTICLE{2020A&A...636A.104B,
	author = {{Belczynski}, K. and {Klencki}, J. and {Fields}, C.~E. and {Olejak}, A. and {Berti}, E. and {Meynet}, G. and {Fryer}, C.~L. and {Holz}, D.~E. and {O'Shaughnessy}, R. and {Brown}, D.~A. and {Bulik}, T. and {Leung}, S.~C. and {Nomoto}, K. and {Madau}, P. and {Hirschi}, R. and {Kaiser}, E. and {Jones}, S. and {Mondal}, S. and {Chruslinska}, M. and {Drozda}, P. and {Gerosa}, D. and {Doctor}, Z. and {Giersz}, M. and {Ekstrom}, S. and {Georgy}, C. and {Askar}, A. and {Baibhav}, V. and {Wysocki}, D. and {Natan}, T. and {Farr}, W.~M. and {Wiktorowicz}, G. and {Coleman Miller}, M. and {Farr}, B. and {Lasota}, J.-P.},
	title = "{Evolutionary roads leading to low effective spins, high black hole masses, and O1/O2 rates for LIGO/Virgo binary black holes}",
	journal = {\aap},
	year = 2020,
	month = apr,
	volume = {636},
	eid = {A104},
	pages = {A104},
	doi = {10.1051/0004-6361/201936528},
	archivePrefix = {arXiv},
	eprint = {1706.07053},
	primaryClass = {astro-ph.HE},
	adsurl = {https://ui.adsabs.harvard.edu/abs/2020A&A...636A.104B}
}

@ARTICLE{2021MNRAS.501.4514C,
	author = {{Costa}, Guglielmo and {Bressan}, Alessandro and {Mapelli}, Michela and {Marigo}, Paola and {Iorio}, Giuliano and {Spera}, Mario},
	title = "{Formation of GW190521 from stellar evolution: the impact of the hydrogen-rich envelope, dredge-up, and $^{12}$C({\ensuremath{\alpha}}, {\ensuremath{\gamma}})$^{16}$O rate on the pair-instability black hole mass gap}",
	journal = {\mnras},
	year = 2021,
	month = mar,
	volume = {501},
	number = {3},
	pages = {4514-4533},
	doi = {10.1093/mnras/staa3916},
	archivePrefix = {arXiv},
	eprint = {2010.02242},
	primaryClass = {astro-ph.SR},
	adsurl = {https://ui.adsabs.harvard.edu/abs/2021MNRAS.501.4514C}
}

@ARTICLE{2020MNRAS.497.1043D,
	author = {{Di Carlo}, Ugo N. and {Mapelli}, Michela and {Bouffanais}, Yann and {Giacobbo}, Nicola and {Santoliquido}, Filippo and {Bressan}, Alessandro and {Spera}, Mario and {Haardt}, Francesco},
	title = "{Binary black holes in the pair instability mass gap}",
	journal = {\mnras},
	year = 2020,
	month = sep,
	volume = {497},
	number = {1},
	pages = {1043-1049},
	doi = {10.1093/mnras/staa1997},
	archivePrefix = {arXiv},
	eprint = {1911.01434},
	primaryClass = {astro-ph.HE},
	adsurl = {https://ui.adsabs.harvard.edu/abs/2020MNRAS.497.1043D}
}

@ARTICLE{2019PhRvD.100d3027R,
	author = {{Rodriguez}, Carl L. and {Zevin}, Michael and {Amaro-Seoane}, Pau and {Chatterjee}, Sourav and {Kremer}, Kyle and {Rasio}, Frederic A. and {Ye}, Claire S.},
	title = "{Black holes: The next generation{\textemdash}repeated mergers in dense star clusters and their gravitational-wave properties}",
	journal = {\prd},
	year = 2019,
	month = aug,
	volume = {100},
	number = {4},
	eid = {043027},
	pages = {043027},
	doi = {10.1103/PhysRevD.100.043027},
	archivePrefix = {arXiv},
	eprint = {1906.10260},
	primaryClass = {astro-ph.HE},
	adsurl = {https://ui.adsabs.harvard.edu/abs/2019PhRvD.100d3027R}
}

@ARTICLE{2019MNRAS.486.5008A,
	author = {{Antonini}, Fabio and {Gieles}, Mark and {Gualandris}, Alessia},
	title = "{Black hole growth through hierarchical black hole mergers in dense star clusters: implications for gravitational wave detections}",
	journal = {\mnras},
	year = 2019,
	month = jul,
	volume = {486},
	number = {4},
	pages = {5008-5021},
	doi = {10.1093/mnras/stz1149},
	archivePrefix = {arXiv},
	eprint = {1811.03640},
	primaryClass = {astro-ph.HE},
	adsurl = {https://ui.adsabs.harvard.edu/abs/2019MNRAS.486.5008A}
}

@ARTICLE{2020PhRvL.125j1102A,
	author = {{Abbott}, B.~P. and {Abbott}, R. and {Abbott}, T.~D. and {LIGO Scientific Collaboration} and {Virgo Collaboration}},
	title = "{GW190521: A Binary Black Hole Merger with a Total Mass of 150  M{\ensuremath{\odot}}}",
	journal = {\prl},
	year = 2020,
	month = sep,
	volume = {125},
	number = {10},
	eid = {101102},
	pages = {101102},
	doi = {10.1103/PhysRevLett.125.101102},
	archivePrefix = {arXiv},
	eprint = {2009.01075},
	primaryClass = {gr-qc},
	adsurl = {https://ui.adsabs.harvard.edu/abs/2020PhRvL.125j1102A}
}

@ARTICLE{2016A&A...594A..97B,
	author = {{Belczynski}, K. and {Heger}, A. and {Gladysz}, W. and {Ruiter}, A.~J. and {Woosley}, S. and {Wiktorowicz}, G. and {Chen}, H.-Y. and {Bulik}, T. and {O'Shaughnessy}, R. and {Holz}, D.~E. and {Fryer}, C.~L. and {Berti}, E.},
	title = "{The effect of pair-instability mass loss on black-hole mergers}",
	journal = {\aap},
	year = 2016,
	month = oct,
	volume = {594},
	eid = {A97},
	pages = {A97},
	doi = {10.1051/0004-6361/201628980},
	archivePrefix = {arXiv},
	eprint = {1607.03116},
	primaryClass = {astro-ph.HE},
	adsurl = {https://ui.adsabs.harvard.edu/abs/2016A&A...594A..97B}
}

@ARTICLE{2017ApJ...836..244W,
	author = {{Woosley}, S.~E.},
	title = "{Pulsational Pair-instability Supernovae}",
	journal = {\apj},
	year = 2017,
	month = feb,
	volume = {836},
	number = {2},
	eid = {244},
	pages = {244},
	doi = {10.3847/1538-4357/836/2/244},
	archivePrefix = {arXiv},
	eprint = {1608.08939},
	primaryClass = {astro-ph.HE},
	adsurl = {https://ui.adsabs.harvard.edu/abs/2017ApJ...836..244W}
}

@ARTICLE{2023Natur.618..712X,
	author = {{Xing}, Qian-Fan and {Zhao}, Gang and {Liu}, Zheng-Wei and {Heger}, Alexander and {Han}, Zhan-Wen and {Aoki}, Wako and {Chen}, Yu-Qin and {Ishigaki}, Miho N. and {Li}, Hai-Ning and {Zhao}, Jing-Kun},
	title = "{A metal-poor star with abundances from a pair-instability supernova}",
	journal = {\nat},
	year = 2023,
	month = jun,
	volume = {618},
	number = {7966},
	pages = {712-715},
	doi = {10.1038/s41586-023-06028-1},
	adsurl = {https://ui.adsabs.harvard.edu/abs/2023Natur.618..712X}
}

@ARTICLE{2020ApJ...897..100V,
	author = {{van Son}, L.~A.~C. and {De Mink}, S.~E. and {Broekgaarden}, F.~S. and {Renzo}, M. and {Justham}, S. and {Laplace}, E. and {Mor{\'a}n-Fraile}, J. and {Hendriks}, D.~D. and {Farmer}, R.},
	title = "{Polluting the Pair-instability Mass Gap for Binary Black Holes through Super-Eddington Accretion in Isolated Binaries}",
	journal = {\apj},
	year = 2020,
	month = jul,
	volume = {897},
	number = {1},
	eid = {100},
	pages = {100},
	doi = {10.3847/1538-4357/ab9809},
	archivePrefix = {arXiv},
	eprint = {2004.05187},
	primaryClass = {astro-ph.HE},
	adsurl = {https://ui.adsabs.harvard.edu/abs/2020ApJ...897..100V}
}

@ARTICLE{2016Natur.534..512B,
	author = {{Belczynski}, Krzysztof and {Holz}, Daniel E. and {Bulik}, Tomasz and {O'Shaughnessy}, Richard},
	title = "{The first gravitational-wave source from the isolated evolution of two stars in the 40-100 solar mass range}",
	journal = {\nat},
	year = 2016,
	month = jun,
	volume = {534},
	number = {7608},
	pages = {512-515},
	doi = {10.1038/nature18322},
	archivePrefix = {arXiv},
	eprint = {1602.04531},
	primaryClass = {astro-ph.HE},
	adsurl = {https://ui.adsabs.harvard.edu/abs/2016Natur.534..512B}
}

@ARTICLE{2016A&A...588A..50M,
	author = {{Marchant}, Pablo and {Langer}, Norbert and {Podsiadlowski}, Philipp and {Tauris}, Thomas M. and {Moriya}, Takashi J.},
	title = "{A new route towards merging massive black holes}",
	journal = {\aap},
	year = 2016,
	month = apr,
	volume = {588},
	eid = {A50},
	pages = {A50},
	doi = {10.1051/0004-6361/201628133},
	archivePrefix = {arXiv},
	eprint = {1601.03718},
	primaryClass = {astro-ph.SR},
	adsurl = {https://ui.adsabs.harvard.edu/abs/2016A&A...588A..50M}
}

@ARTICLE{2021MNRAS.504..146V,
	author = {{Vink}, Jorick S. and {Higgins}, Erin R. and {Sander}, Andreas A.~C. and {Sabhahit}, Gautham N.},
	title = "{Maximum black hole mass across cosmic time}",
	journal = {\mnras},
	year = 2021,
	month = jun,
	volume = {504},
	number = {1},
	pages = {146-154},
	doi = {10.1093/mnras/stab842},
	archivePrefix = {arXiv},
	eprint = {2010.11730},
	primaryClass = {astro-ph.HE},
	adsurl = {https://ui.adsabs.harvard.edu/abs/2021MNRAS.504..146V}
}

@ARTICLE{2023MNRAS.519.5191B,
	author = {{Ballone}, Alessandro and {Costa}, Guglielmo and {Mapelli}, Michela and {MacLeod}, Morgan and {Torniamenti}, Stefano and {Pacheco-Arias}, Juan Manuel},
	title = "{Formation of black holes in the pair-instability mass gap: hydrodynamical simulations of a head-on massive star collision}",
	journal = {\mnras},
	year = 2023,
	month = mar,
	volume = {519},
	number = {4},
	pages = {5191-5201},
	doi = {10.1093/mnras/stac3752},
	archivePrefix = {arXiv},
	eprint = {2204.03493},
	primaryClass = {astro-ph.SR},
	adsurl = {https://ui.adsabs.harvard.edu/abs/2023MNRAS.519.5191B}
}

@ARTICLE{1988ApJ...329..764L,
	author = {{Livio}, Mario and {Soker}, Noam},
	title = "{The Common Envelope Phase in the Evolution of Binary Stars}",
	journal = {\apj},
	year = 1988,
	month = jun,
	volume = {329},
	pages = {764},
	doi = {10.1086/166419},
	adsurl = {https://ui.adsabs.harvard.edu/abs/1988ApJ...329..764L}
}

@article{Patton_2025,
	doi = {10.3847/1538-4357/addd0f},
	url = {https://doi.org/10.3847/1538-4357/addd0f},
	year = {2025},
	month = {jul},
	publisher = {The American Astronomical Society},
	volume = {987},
	number = {2},
	pages = {212},
	author = {Patton, Rachel A. and Pinsonneault, Marc H. and Thompson, Todd A.},
	title = {The Structure and Evolution of a High-mass Stellar Merger in the Hertzsprung Gap},
	journal = {The Astrophysical Journal},
}

@ARTICLE{2022MNRAS.511.5797M,
	author = {{Mapelli}, Michela and {Bouffanais}, Yann and {Santoliquido}, Filippo and {Arca Sedda}, Manuel and {Artale}, M. Celeste},
	title = "{The cosmic evolution of binary black holes in young, globular, and nuclear star clusters: rates, masses, spins, and mixing fractions}",
	journal = {\mnras},
	year = 2022,
	month = apr,
	volume = {511},
	number = {4},
	pages = {5797-5816},
	doi = {10.1093/mnras/stac422},
	archivePrefix = {arXiv},
	eprint = {2109.06222},
	primaryClass = {astro-ph.HE},
	adsurl = {https://ui.adsabs.harvard.edu/abs/2022MNRAS.511.5797M}
}

@ARTICLE{2014MNRAS.441.3703Z,
	author = {{Ziosi}, Brunetto Marco and {Mapelli}, Michela and {Branchesi}, Marica and {Tormen}, Giuseppe},
	title = "{Dynamics of stellar black holes in young star clusters with different metallicities - II. Black hole-black hole binaries}",
	journal = {\mnras},
	year = 2014,
	month = jul,
	volume = {441},
	number = {4},
	pages = {3703-3717},
	doi = {10.1093/mnras/stu824},
	archivePrefix = {arXiv},
	eprint = {1404.7147},
	primaryClass = {astro-ph.GA},
	adsurl = {https://ui.adsabs.harvard.edu/abs/2014MNRAS.441.3703Z}
}

@ARTICLE{2019MNRAS.487.2947D,
	author = {{Di Carlo}, Ugo N. and {Giacobbo}, Nicola and {Mapelli}, Michela and {Pasquato}, Mario and {Spera}, Mario and {Wang}, Long and {Haardt}, Francesco},
	title = "{Merging black holes in young star clusters}",
	journal = {\mnras},
	year = 2019,
	month = aug,
	volume = {487},
	number = {2},
	pages = {2947-2960},
	doi = {10.1093/mnras/stz1453},
	archivePrefix = {arXiv},
	eprint = {1901.00863},
	primaryClass = {astro-ph.HE},
	adsurl = {https://ui.adsabs.harvard.edu/abs/2019MNRAS.487.2947D}
}

@ARTICLE{2021NatAs...5..749G,
	author = {{Gerosa}, Davide and {Fishbach}, Maya},
	title = "{Hierarchical mergers of stellar-mass black holes and their gravitational-wave signatures}",
	journal = {Nature Astronomy},
	year = 2021,
	month = jul,
	volume = {5},
	pages = {749-760},
	doi = {10.1038/s41550-021-01398-w},
	archivePrefix = {arXiv},
	eprint = {2105.03439},
	primaryClass = {astro-ph.HE},
	adsurl = {https://ui.adsabs.harvard.edu/abs/2021NatAs...5..749G}
}

@ARTICLE{2018PhRvL.121p1103F,
	author = {{Fragione}, Giacomo and {Kocsis}, Bence},
	title = "{Black Hole Mergers from an Evolving Population of Globular Clusters}",
	journal = {\prl},
	year = 2018,
	month = oct,
	volume = {121},
	number = {16},
	eid = {161103},
	pages = {161103},
	doi = {10.1103/PhysRevLett.121.161103},
	archivePrefix = {arXiv},
	eprint = {1806.02351},
	primaryClass = {astro-ph.GA},
	adsurl = {https://ui.adsabs.harvard.edu/abs/2018PhRvL.121p1103F}
}

@ARTICLE{2000ApJ...528L..17P,
	author = {{Portegies Zwart}, Simon F. and {McMillan}, Stephen L.~W.},
	title = "{Black Hole Mergers in the Universe}",
	journal = {\apjl},
	year = 2000,
	month = jan,
	volume = {528},
	number = {1},
	pages = {L17-L20},
	doi = {10.1086/312422},
	archivePrefix = {arXiv},
	eprint = {astro-ph/9910061},
	primaryClass = {astro-ph},
	adsurl = {https://ui.adsabs.harvard.edu/abs/2000ApJ...528L..17P}
}

@ARTICLE{2020ApJ...893...35D,
	author = {{Doctor}, Z. and {Wysocki}, D. and {O'Shaughnessy}, R. and {Holz}, D.~E. and {Farr}, B.},
	title = "{Black Hole Coagulation: Modeling Hierarchical Mergers in Black Hole Populations}",
	journal = {\apj},
	year = 2020,
	month = apr,
	volume = {893},
	number = {1},
	eid = {35},
	pages = {35},
	doi = {10.3847/1538-4357/ab7fac},
	archivePrefix = {arXiv},
	eprint = {1911.04424},
	primaryClass = {astro-ph.HE},
	adsurl = {https://ui.adsabs.harvard.edu/abs/2020ApJ...893...35D}
}

@ARTICLE{2020ApJ...903...45K,
	author = {{Kremer}, Kyle and {Spera}, Mario and {Becker}, Devin and {Chatterjee}, Sourav and {Di Carlo}, Ugo N. and {Fragione}, Giacomo and {Rodriguez}, Carl L. and {Ye}, Claire S. and {Rasio}, Frederic A.},
	title = "{Populating the Upper Black Hole Mass Gap through Stellar Collisions in Young Star Clusters}",
	journal = {\apj},
	year = 2020,
	month = nov,
	volume = {903},
	number = {1},
	eid = {45},
	pages = {45},
	doi = {10.3847/1538-4357/abb945},
	archivePrefix = {arXiv},
	eprint = {2006.10771},
	primaryClass = {astro-ph.HE},
	adsurl = {https://ui.adsabs.harvard.edu/abs/2020ApJ...903...45K}
}

@ARTICLE{2021MNRAS.505..339M,
	author = {{Mapelli}, Michela and {Dall'Amico}, Marco and {Bouffanais}, Yann and {Giacobbo}, Nicola and {Arca Sedda}, Manuel and {Artale}, M. Celeste and {Ballone}, Alessandro and {Di Carlo}, Ugo N. and {Iorio}, Giuliano and {Santoliquido}, Filippo and {Torniamenti}, Stefano},
	title = "{Hierarchical black hole mergers in young, globular and nuclear star clusters: the effect of metallicity, spin and cluster properties}",
	journal = {\mnras},
	year = 2021,
	month = jul,
	volume = {505},
	number = {1},
	pages = {339-358},
	doi = {10.1093/mnras/stab1334},
	archivePrefix = {arXiv},
	eprint = {2103.05016},
	primaryClass = {astro-ph.HE},
	adsurl = {https://ui.adsabs.harvard.edu/abs/2021MNRAS.505..339M}
}

@ARTICLE{2017ApJ...835..165B,
	author = {{Bartos}, Imre and {Kocsis}, Bence and {Haiman}, Zolt{\'a}n and {M{\'a}rka}, Szabolcs},
	title = "{Rapid and Bright Stellar-mass Binary Black Hole Mergers in Active Galactic Nuclei}",
	journal = {\apj},
	year = 2017,
	month = feb,
	volume = {835},
	number = {2},
	eid = {165},
	pages = {165},
	doi = {10.3847/1538-4357/835/2/165},
	archivePrefix = {arXiv},
	eprint = {1602.03831},
	primaryClass = {astro-ph.HE},
	adsurl = {https://ui.adsabs.harvard.edu/abs/2017ApJ...835..165B}
}

@ARTICLE{2017MNRAS.464..946S,
	author = {{Stone}, Nicholas C. and {Metzger}, Brian D. and {Haiman}, Zolt{\'a}n},
	title = "{Assisted inspirals of stellar mass black holes embedded in AGN discs: solving the `final au problem'}",
	journal = {\mnras},
	year = 2017,
	month = jan,
	volume = {464},
	number = {1},
	pages = {946-954},
	doi = {10.1093/mnras/stw2260},
	archivePrefix = {arXiv},
	eprint = {1602.04226},
	primaryClass = {astro-ph.GA},
	adsurl = {https://ui.adsabs.harvard.edu/abs/2017MNRAS.464..946S}
}

@ARTICLE{2016ApJ...819L..17B,
	author = {{Bellovary}, Jillian M. and {Mac Low}, Mordecai-Mark and {McKernan}, Barry and {Ford}, K.~E. Saavik},
	title = "{Migration Traps in Disks around Supermassive Black Holes}",
	journal = {\apjl},
	year = 2016,
	month = mar,
	volume = {819},
	number = {2},
	eid = {L17},
	pages = {L17},
	doi = {10.3847/2041-8205/819/2/L17},
	archivePrefix = {arXiv},
	eprint = {1511.00005},
	primaryClass = {astro-ph.GA},
	adsurl = {https://ui.adsabs.harvard.edu/abs/2016ApJ...819L..17B}
}

@ARTICLE{2018ApJ...866...66M,
	author = {{McKernan}, Barry and {Ford}, K.~E. Saavik and {Bellovary}, J. and {Leigh}, N.~W.~C. and {Haiman}, Z. and {Kocsis}, B. and {Lyra}, W. and {Mac Low}, M.-M. and {Metzger}, B. and {O'Dowd}, M. and {Endlich}, S. and {Rosen}, D.~J.},
	title = "{Constraining Stellar-mass Black Hole Mergers in AGN Disks Detectable with LIGO}",
	journal = {\apj},
	year = 2018,
	month = oct,
	volume = {866},
	number = {1},
	eid = {66},
	pages = {66},
	doi = {10.3847/1538-4357/aadae5},
	archivePrefix = {arXiv},
	eprint = {1702.07818},
	primaryClass = {astro-ph.HE},
	adsurl = {https://ui.adsabs.harvard.edu/abs/2018ApJ...866...66M}
}

@ARTICLE{2020ApJ...898...25T,
	author = {{Tagawa}, Hiromichi and {Haiman}, Zolt{\'a}n and {Kocsis}, Bence},
	title = "{Formation and Evolution of Compact-object Binaries in AGN Disks}",
	journal = {\apj},
	year = 2020,
	month = jul,
	volume = {898},
	number = {1},
	eid = {25},
	pages = {25},
	doi = {10.3847/1538-4357/ab9b8c},
	archivePrefix = {arXiv},
	eprint = {1912.08218},
	primaryClass = {astro-ph.GA},
	adsurl = {https://ui.adsabs.harvard.edu/abs/2020ApJ...898...25T}
}

@ARTICLE{2019PhRvL.123r1101Y,
	author = {{Yang}, Y. and {Bartos}, I. and {Gayathri}, V. and {Ford}, K.~E.~S. and {Haiman}, Z. and {Klimenko}, S. and {Kocsis}, B. and {M{\'a}rka}, S. and {M{\'a}rka}, Z. and {McKernan}, B. and {O'Shaughnessy}, R.},
	title = "{Hierarchical Black Hole Mergers in Active Galactic Nuclei}",
	journal = {\prl},
	year = 2019,
	month = nov,
	volume = {123},
	number = {18},
	eid = {181101},
	pages = {181101},
	doi = {10.1103/PhysRevLett.123.181101},
	archivePrefix = {arXiv},
	eprint = {1906.09281},
	primaryClass = {astro-ph.HE},
	adsurl = {https://ui.adsabs.harvard.edu/abs/2019PhRvL.123r1101Y}
}

@ARTICLE{2021ApJ...908..194T,
	author = {{Tagawa}, Hiromichi and {Kocsis}, Bence and {Haiman}, Zolt{\'a}n and {Bartos}, Imre and {Omukai}, Kazuyuki and {Samsing}, Johan},
	title = "{Mass-gap Mergers in Active Galactic Nuclei}",
	journal = {\apj},
	year = 2021,
	month = feb,
	volume = {908},
	number = {2},
	eid = {194},
	pages = {194},
	doi = {10.3847/1538-4357/abd555},
	archivePrefix = {arXiv},
	eprint = {2012.00011},
	primaryClass = {astro-ph.HE},
	adsurl = {https://ui.adsabs.harvard.edu/abs/2021ApJ...908..194T}
}

@ARTICLE{2014ApJS..215...15S,
	author = {{Sana}, H. and {Le Bouquin}, J.-B. and {Lacour}, S. and {Berger}, J.-P. and {Duvert}, G. and {Gauchet}, L. and {Norris}, B. and {Olofsson}, J. and {Pickel}, D. and {Zins}, G. and {Absil}, O. and {de Koter}, A. and {Kratter}, K. and {Schnurr}, O. and {Zinnecker}, H.},
	title = "{Southern Massive Stars at High Angular Resolution: Observational Campaign and Companion Detection}",
	journal = {\apjs},
	year = 2014,
	month = nov,
	volume = {215},
	number = {1},
	eid = {15},
	pages = {15},
	doi = {10.1088/0067-0049/215/1/15},
	archivePrefix = {arXiv},
	eprint = {1409.6304},
	primaryClass = {astro-ph.SR},
	adsurl = {https://ui.adsabs.harvard.edu/abs/2014ApJS..215...15S}
}

@ARTICLE{2019ApJ...878...85S,
	author = {{Secunda}, Amy and {Bellovary}, Jillian and {Mac Low}, Mordecai-Mark and {Ford}, K.~E. Saavik and {McKernan}, Barry and {Leigh}, Nathan W.~C. and {Lyra}, Wladimir and {S{\'a}ndor}, Zsolt},
	title = "{Orbital Migration of Interacting Stellar Mass Black Holes in Disks around Supermassive Black Holes}",
	journal = {\apj},
	year = 2019,
	month = jun,
	volume = {878},
	number = {2},
	eid = {85},
	pages = {85},
	doi = {10.3847/1538-4357/ab20ca},
	archivePrefix = {arXiv},
	eprint = {1807.02859},
	primaryClass = {astro-ph.HE},
	adsurl = {https://ui.adsabs.harvard.edu/abs/2019ApJ...878...85S}
}

@ARTICLE{2020A&A...640A..56R,
	author = {{Renzo}, M. and {Farmer}, R. and {Justham}, S. and {G{\"o}tberg}, Y. and {de Mink}, S.~E. and {Zapartas}, E. and {Marchant}, P. and {Smith}, N.},
	title = "{Predictions for the hydrogen-free ejecta of pulsational pair-instability supernovae}",
	journal = {\aap},
	year = 2020,
	month = aug,
	volume = {640},
	eid = {A56},
	pages = {A56},
	doi = {10.1051/0004-6361/202037710},
	archivePrefix = {arXiv},
	eprint = {2002.05077},
	primaryClass = {astro-ph.SR},
	adsurl = {https://ui.adsabs.harvard.edu/abs/2020A&A...640A..56R}
}

@ARTICLE{2017MNRAS.470.4739S,
	author = {{Spera}, Mario and {Mapelli}, Michela},
	title = "{Very massive stars, pair-instability supernovae and intermediate-mass black holes with the sevn code}",
	journal = {\mnras},
	year = 2017,
	month = oct,
	volume = {470},
	number = {4},
	pages = {4739-4749},
	doi = {10.1093/mnras/stx1576},
	archivePrefix = {arXiv},
	eprint = {1706.06109},
	primaryClass = {astro-ph.SR},
	adsurl = {https://ui.adsabs.harvard.edu/abs/2017MNRAS.470.4739S}
}

@ARTICLE{2019ApJ...882...36M,
	author = {{Marchant}, Pablo and {Renzo}, Mathieu and {Farmer}, Robert and {Pappas}, Kaliroe M.~W. and {Taam}, Ronald E. and {de Mink}, Selma E. and {Kalogera}, Vassiliki},
	title = "{Pulsational Pair-instability Supernovae in Very Close Binaries}",
	journal = {\apj},
	year = 2019,
	month = sep,
	volume = {882},
	number = {1},
	eid = {36},
	pages = {36},
	doi = {10.3847/1538-4357/ab3426},
	archivePrefix = {arXiv},
	eprint = {1810.13412},
	primaryClass = {astro-ph.HE},
	adsurl = {https://ui.adsabs.harvard.edu/abs/2019ApJ...882...36M}
}

@ARTICLE{2012Sci...337..444S,
	author = {{Sana}, H. and {de Mink}, S.~E. and {de Koter}, A. and {Langer}, N. and {Evans}, C.~J. and {Gieles}, M. and {Gosset}, E. and {Izzard}, R.~G. and {Le Bouquin}, J.-B. and {Schneider}, F.~R.~N.},
	title = "{Binary Interaction Dominates the Evolution of Massive Stars}",
	journal = {Science},
	year = 2012,
	month = jul,
	volume = {337},
	number = {6093},
	pages = {444},
	doi = {10.1126/science.1223344},
	archivePrefix = {arXiv},
	eprint = {1207.6397},
	primaryClass = {astro-ph.SR},
	adsurl = {https://ui.adsabs.harvard.edu/abs/2012Sci...337..444S}
}

@ARTICLE{2013ARA&A..51..269D,
	author = {{Duch{\^e}ne}, Gaspard and {Kraus}, Adam},
	title = "{Stellar Multiplicity}",
	journal = {\araa},
	year = 2013,
	month = aug,
	volume = {51},
	number = {1},
	pages = {269-310},
	doi = {10.1146/annurev-astro-081710-102602},
	archivePrefix = {arXiv},
	eprint = {1303.3028},
	primaryClass = {astro-ph.SR},
	adsurl = {https://ui.adsabs.harvard.edu/abs/2013ARA&A..51..269D}
}

@ARTICLE{2014ApJS..213...34K,
	author = {{Kobulnicky}, Henry A. and {Kiminki}, Daniel C. and {Lundquist}, Michael J. and {Burke}, Jamison and {Chapman}, James and {Keller}, Erica and {Lester}, Kathryn and {Rolen}, Emily K. and {Topel}, Eric and {Bhattacharjee}, Anirban and {Smullen}, Rachel A. and {Vargas {\'A}lvarez}, Carlos A. and {Runnoe}, Jessie C. and {Dale}, Daniel A. and {Brotherton}, Michael M.},
	title = "{Toward Complete Statistics of Massive Binary Stars: Penultimate Results from the Cygnus OB2 Radial Velocity Survey}",
	journal = {\apjs},
	year = 2014,
	month = aug,
	volume = {213},
	number = {2},
	eid = {34},
	pages = {34},
	doi = {10.1088/0067-0049/213/2/34},
	archivePrefix = {arXiv},
	eprint = {1406.6655},
	primaryClass = {astro-ph.SR},
	adsurl = {https://ui.adsabs.harvard.edu/abs/2014ApJS..213...34K}
}

@ARTICLE{2017ApJ...841...77A,
	author = {{Antonini}, Fabio and {Toonen}, Silvia and {Hamers}, Adrian S.},
	title = "{Binary Black Hole Mergers from Field Triples: Properties, Rates, and the Impact of Stellar Evolution}",
	journal = {\apj},
	year = 2017,
	month = jun,
	volume = {841},
	number = {2},
	eid = {77},
	pages = {77},
	doi = {10.3847/1538-4357/aa6f5e},
	archivePrefix = {arXiv},
	eprint = {1703.06614},
	primaryClass = {astro-ph.GA},
	adsurl = {https://ui.adsabs.harvard.edu/abs/2017ApJ...841...77A}
}

@ARTICLE{2019MNRAS.486.4443F,
	author = {{Fragione}, Giacomo and {Loeb}, Abraham},
	title = "{Black hole-neutron star mergers from triples}",
	journal = {\mnras},
	year = 2019,
	month = jul,
	volume = {486},
	number = {3},
	pages = {4443-4450},
	doi = {10.1093/mnras/stz1131},
	archivePrefix = {arXiv},
	eprint = {1903.10511},
	primaryClass = {astro-ph.GA},
	adsurl = {https://ui.adsabs.harvard.edu/abs/2019MNRAS.486.4443F}
}

@ARTICLE{2021ApJ...907L..19V,
	author = {{Vigna-G{\'o}mez}, Alejandro and {Toonen}, Silvia and {Ramirez-Ruiz}, Enrico and {Leigh}, Nathan W.~C. and {Riley}, Jeff and {Haster}, Carl-Johan},
	title = "{Massive Stellar Triples Leading to Sequential Binary Black Hole Mergers in the Field}",
	journal = {\apjl},
	year = 2021,
	month = jan,
	volume = {907},
	number = {1},
	eid = {L19},
	pages = {L19},
	doi = {10.3847/2041-8213/abd5b7},
	archivePrefix = {arXiv},
	eprint = {2010.13669},
	primaryClass = {astro-ph.HE},
	adsurl = {https://ui.adsabs.harvard.edu/abs/2021ApJ...907L..19V}
}

@ARTICLE{2022A&A...661A..61T,
	author = {{Toonen}, S. and {Boekholt}, T.~C.~N. and {Portegies Zwart}, S.},
	title = "{Stellar triples on the edge. Comprehensive overview of the evolution of destabilised triples leading to stellar and binary exotica}",
	journal = {\aap},
	year = 2022,
	month = may,
	volume = {661},
	eid = {A61},
	pages = {A61},
	doi = {10.1051/0004-6361/202141991},
	archivePrefix = {arXiv},
	eprint = {2108.04272},
	primaryClass = {astro-ph.SR},
	adsurl = {https://ui.adsabs.harvard.edu/abs/2022A&A...661A..61T}
}

@ARTICLE{2018ApJ...863...68L,
	author = {{Liu}, Bin and {Lai}, Dong},
	title = "{Black Hole and Neutron Star Binary Mergers in Triple Systems: Merger Fraction and Spin-Orbit Misalignment}",
	journal = {\apj},
	year = 2018,
	month = aug,
	volume = {863},
	number = {1},
	eid = {68},
	pages = {68},
	doi = {10.3847/1538-4357/aad09f},
	archivePrefix = {arXiv},
	eprint = {1805.03202},
	primaryClass = {astro-ph.HE},
	adsurl = {https://ui.adsabs.harvard.edu/abs/2018ApJ...863...68L}
}

@ARTICLE{2002ApJ...565..385U,
	author = {{Umeda}, Hideyuki and {Nomoto}, Ken'ichi},
	title = "{Nucleosynthesis of Zinc and Iron Peak Elements in Population III Type II Supernovae: Comparison with Abundances of Very Metal Poor Halo Stars}",
	journal = {\apj},
	year = 2002,
	month = jan,
	volume = {565},
	number = {1},
	pages = {385-404},
	doi = {10.1086/323946},
	archivePrefix = {arXiv},
	eprint = {astro-ph/0103241},
	primaryClass = {astro-ph},
	adsurl = {https://ui.adsabs.harvard.edu/abs/2002ApJ...565..385U}
}

@ARTICLE{2014Sci...345..912A,
	author = {{Aoki}, W. and {Tominaga}, N. and {Beers}, T.~C. and {Honda}, S. and {Lee}, Y.~S.},
	title = "{A chemical signature of first-generation very massive stars}",
	journal = {Science},
	year = 2014,
	month = aug,
	volume = {345},
	number = {6199},
	pages = {912-915},
	doi = {10.1126/science.1252633},
	adsurl = {https://ui.adsabs.harvard.edu/abs/2014Sci...345..912A}
}

@ARTICLE{2002Sci...295...93A,
	author = {{Abel}, Tom and {Bryan}, Greg L. and {Norman}, Michael L.},
	title = "{The Formation of the First Star in the Universe}",
	journal = {Science},
	year = 2002,
	month = jan,
	volume = {295},
	number = {5552},
	pages = {93-98},
	doi = {10.1126/science.1063991},
	archivePrefix = {arXiv},
	eprint = {astro-ph/0112088},
	primaryClass = {astro-ph},
	adsurl = {https://ui.adsabs.harvard.edu/abs/2002Sci...295...93A}
}

@ARTICLE{2004ARA&A..42...79B,
	author = {{Bromm}, Volker and {Larson}, Richard B.},
	title = "{The First Stars}",
	journal = {\araa},
	year = 2004,
	month = sep,
	volume = {42},
	number = {1},
	pages = {79-118},
	doi = {10.1146/annurev.astro.42.053102.134034},
	archivePrefix = {arXiv},
	eprint = {astro-ph/0311019},
	primaryClass = {astro-ph},
	adsurl = {https://ui.adsabs.harvard.edu/abs/2004ARA&A..42...79B}
}

@ARTICLE{2014ApJ...781...60H,
	author = {{Hirano}, Shingo and {Hosokawa}, Takashi and {Yoshida}, Naoki and {Umeda}, Hideyuki and {Omukai}, Kazuyuki and {Chiaki}, Gen and {Yorke}, Harold W.},
	title = "{One Hundred First Stars: Protostellar Evolution and the Final Masses}",
	journal = {\apj},
	year = 2014,
	month = feb,
	volume = {781},
	number = {2},
	eid = {60},
	pages = {60},
	doi = {10.1088/0004-637X/781/2/60},
	archivePrefix = {arXiv},
	eprint = {1308.4456},
	primaryClass = {astro-ph.CO},
	adsurl = {https://ui.adsabs.harvard.edu/abs/2014ApJ...781...60H}
}

@ARTICLE{2018ApJ...857..111T,
	author = {{Takahashi}, Koh and {Yoshida}, Takashi and {Umeda}, Hideyuki},
	title = "{Stellar Yields of Rotating First Stars. II. Pair-instability Supernovae and Comparison with Observations}",
	journal = {\apj},
	year = 2018,
	month = apr,
	volume = {857},
	number = {2},
	eid = {111},
	pages = {111},
	doi = {10.3847/1538-4357/aab95f},
	archivePrefix = {arXiv},
	eprint = {1803.06630},
	primaryClass = {astro-ph.SR},
	adsurl = {https://ui.adsabs.harvard.edu/abs/2018ApJ...857..111T}
}

@ARTICLE{2010PhRvD..82f4016S,
	author = {{Santamar{\'\i}a}, L. and {Ohme}, F. and {Ajith}, P. and {Br{\"u}gmann}, B. and {Dorband}, N. and {Hannam}, M. and {Husa}, S. and {M{\"o}sta}, P. and {Pollney}, D. and {Reisswig}, C. and {Robinson}, E.~L. and {Seiler}, J. and {Krishnan}, B.},
	title = "{Matching post-Newtonian and numerical relativity waveforms: Systematic errors and a new phenomenological model for nonprecessing black hole binaries}",
	journal = {\prd},
	year = 2010,
	month = sep,
	volume = {82},
	number = {6},
	eid = {064016},
	pages = {064016},
	doi = {10.1103/PhysRevD.82.064016},
	archivePrefix = {arXiv},
	eprint = {1005.3306},
	primaryClass = {gr-qc},
	adsurl = {https://ui.adsabs.harvard.edu/abs/2010PhRvD..82f4016S}
}

@ARTICLE{2017ApJS..230...15M,
	author = {{Moe}, Maxwell and {Di Stefano}, Rosanne},
	title = "{Mind Your Ps and Qs: The Interrelation between Period (P) and Mass-ratio (Q) Distributions of Binary Stars}",
	journal = {\apjs},
	year = 2017,
	month = jun,
	volume = {230},
	number = {2},
	eid = {15},
	pages = {15},
	doi = {10.3847/1538-4365/aa6fb6},
	archivePrefix = {arXiv},
	eprint = {1606.05347},
	primaryClass = {astro-ph.SR},
	adsurl = {https://ui.adsabs.harvard.edu/abs/2017ApJS..230...15M}
}

@ARTICLE{2022MNRAS.516.1406S,
	author = {{Stegmann}, Jakob and {Antonini}, Fabio and {Moe}, Maxwell},
	title = "{Evolution of massive stellar triples and implications for compact object binary formation}",
	journal = {\mnras},
	year = 2022,
	month = oct,
	volume = {516},
	number = {1},
	pages = {1406-1427},
	doi = {10.1093/mnras/stac2192},
	archivePrefix = {arXiv},
	eprint = {2112.10786},
	primaryClass = {astro-ph.SR},
	adsurl = {https://ui.adsabs.harvard.edu/abs/2022MNRAS.516.1406S}
}

@ARTICLE{2001MNRAS.321..398M,
	author = {{Mardling}, Rosemary A. and {Aarseth}, Sverre J.},
	title = "{Tidal interactions in star cluster simulations}",
	journal = {\mnras},
	year = 2001,
	month = mar,
	volume = {321},
	number = {3},
	pages = {398-420},
	doi = {10.1046/j.1365-8711.2001.03974.x},
	adsurl = {https://ui.adsabs.harvard.edu/abs/2001MNRAS.321..398M}
}

@ARTICLE{2026MNRAS.547ag192D,
       author = {{Di Stefano}, Rosanne and {Khwaja}, Amaan and {Kobayashi}, Chiaki},
        title = "{SCATTER common envelope formalism for triples}",
      journal = {\mnras},
         year = 2026,
        month = apr,
       volume = {547},
       number = {2},
          eid = {stag192},
        pages = {stag192},
          doi = {10.1093/mnras/stag192},
archivePrefix = {arXiv},
       eprint = {2511.04857},
 primaryClass = {astro-ph.SR},
       adsurl = {https://ui.adsabs.harvard.edu/abs/2026MNRAS.547ag192D}
}

@ARTICLE{2013A&ARv..21...59I,
       author = {{Ivanova}, N. and {Justham}, S. and {Chen}, X. and {De Marco}, O. and {Fryer}, C.~L. and {Gaburov}, E. and {Ge}, H. and {Glebbeek}, E. and {Han}, Z. and {Li}, X.-D. and {Lu}, G. and {Marsh}, T. and {Podsiadlowski}, P. and {Potter}, A. and {Soker}, N. and {Taam}, R. and {Tauris}, T.~M. and {van den Heuvel}, E.~P.~J. and {Webbink}, R.~F.},
        title = "{Common envelope evolution: where we stand and how we can move forward}",
      journal = {\aapr},
         year = 2013,
        month = feb,
       volume = {21},
          eid = {59},
        pages = {59},
          doi = {10.1007/s00159-013-0059-2},
archivePrefix = {arXiv},
       eprint = {1209.4302},
 primaryClass = {astro-ph.HE},
       adsurl = {https://ui.adsabs.harvard.edu/abs/2013A&ARv..21...59I}
}

@ARTICLE{2016ComAC...3....6T,
       author = {{Toonen}, Silvia and {Hamers}, Adrian and {Portegies Zwart}, Simon},
        title = "{The evolution of hierarchical triple star-systems}",
      journal = {Computational Astrophysics and Cosmology},
         year = 2016,
        month = dec,
       volume = {3},
       number = {1},
          eid = {6},
        pages = {6},
          doi = {10.1186/s40668-016-0019-0},
archivePrefix = {arXiv},
       eprint = {1612.06172},
 primaryClass = {astro-ph.SR},
       adsurl = {https://ui.adsabs.harvard.edu/abs/2016ComAC...3....6T}
}

@ARTICLE{2023ApJ...944...87D,
       author = {{Di Stefano}, Rosanne and {Kruckow}, Matthias U. and {Gao}, Yan and {Neunteufel}, Patrick G. and {Kobayashi}, Chiaki},
        title = "{SCATTER: A New Common Envelope Formalism}",
      journal = {\apj},
         year = 2023,
        month = feb,
       volume = {944},
       number = {1},
          eid = {87},
        pages = {87},
          doi = {10.3847/1538-4357/acae9b},
archivePrefix = {arXiv},
       eprint = {2212.06770},
 primaryClass = {astro-ph.HE},
       adsurl = {https://ui.adsabs.harvard.edu/abs/2023ApJ...944...87D}
}

@ARTICLE{2025PhRvD.112j3005L,
       author = {{Li}, Lei and {L{\"u}}, Guoliang and {Zhu}, Chunhua and {Guo}, Sufen and {Ge}, Hongwei and {Gu}, Weimin and {Li}, Zhuowen and {He}, Xiaolong},
        title = "{Explanation of the mass distribution of binary black hole mergers}",
      journal = {\prd},
         year = 2025,
        month = nov,
       volume = {112},
       number = {10},
          eid = {103005},
        pages = {103005},
          doi = {10.1103/drq9-dpy4},
archivePrefix = {arXiv},
       eprint = {2510.08231},
 primaryClass = {astro-ph.HE},
       adsurl = {https://ui.adsabs.harvard.edu/abs/2025PhRvD.112j3005L}
}

@ARTICLE{2006A&A...460..199Y,
       author = {{Yoon}, S.-C. and {Langer}, N. and {Norman}, C.},
        title = "{Single star progenitors of long gamma-ray bursts. I. Model grids and redshift dependent GRB rate}",
      journal = {\aap},
         year = 2006,
        month = dec,
       volume = {460},
       number = {1},
        pages = {199-208},
          doi = {10.1051/0004-6361:20065912},
archivePrefix = {arXiv},
       eprint = {astro-ph/0606637},
 primaryClass = {astro-ph},
       adsurl = {https://ui.adsabs.harvard.edu/abs/2006A&A...460..199Y}
}

@ARTICLE{2009A&A...497..243D,
       author = {{de Mink}, S.~E. and {Cantiello}, M. and {Langer}, N. and {Pols}, O.~R. and {Brott}, I. and {Yoon}, S.-Ch.},
        title = "{Rotational mixing in massive binaries. Detached short-period systems}",
      journal = {\aap},
         year = 2009,
        month = apr,
       volume = {497},
       number = {1},
        pages = {243-253},
          doi = {10.1051/0004-6361/200811439},
archivePrefix = {arXiv},
       eprint = {0902.1751},
 primaryClass = {astro-ph.SR},
       adsurl = {https://ui.adsabs.harvard.edu/abs/2009A&A...497..243D}
}

@ARTICLE{2016MNRAS.458.2634M,
       author = {{Mandel}, Ilya and {de Mink}, Selma E.},
        title = "{Merging binary black holes formed through chemically homogeneous evolution in short-period stellar binaries}",
      journal = {\mnras},
         year = 2016,
        month = may,
       volume = {458},
       number = {3},
        pages = {2634-2647},
          doi = {10.1093/mnras/stw379},
archivePrefix = {arXiv},
       eprint = {1601.00007},
 primaryClass = {astro-ph.HE},
       adsurl = {https://ui.adsabs.harvard.edu/abs/2016MNRAS.458.2634M}
}

@ARTICLE{2016MNRAS.460.3545D,
       author = {{de Mink}, S.~E. and {Mandel}, I.},
        title = "{The chemically homogeneous evolutionary channel for binary black hole mergers: rates and properties of gravitational-wave events detectable by advanced LIGO}",
      journal = {\mnras},
         year = 2016,
        month = aug,
       volume = {460},
       number = {4},
        pages = {3545-3553},
          doi = {10.1093/mnras/stw1219},
archivePrefix = {arXiv},
       eprint = {1603.02291},
 primaryClass = {astro-ph.HE},
       adsurl = {https://ui.adsabs.harvard.edu/abs/2016MNRAS.460.3545D}
}

@ARTICLE{2021MNRAS.505..663R,
       author = {{Riley}, Jeff and {Mandel}, Ilya and {Marchant}, Pablo and {Butler}, Ellen and {Nathaniel}, Kaila and {Neijssel}, Coenraad and {Shortt}, Spencer and {Vigna-G{\'o}mez}, Alejandro},
        title = "{Chemically homogeneous evolution: a rapid population synthesis approach}",
      journal = {\mnras},
         year = 2021,
        month = jul,
       volume = {505},
       number = {1},
        pages = {663-676},
          doi = {10.1093/mnras/stab1291},
archivePrefix = {arXiv},
       eprint = {2010.00002},
 primaryClass = {astro-ph.SR},
       adsurl = {https://ui.adsabs.harvard.edu/abs/2021MNRAS.505..663R}
}

@ARTICLE{2002A&A...381..923S,
       author = {{Spruit}, H.~C.},
        title = "{Dynamo action by differential rotation in a stably stratified stellar interior}",
      journal = {\aap},
         year = 2002,
        month = jan,
       volume = {381},
        pages = {923-932},
          doi = {10.1051/0004-6361:20011465},
archivePrefix = {arXiv},
       eprint = {astro-ph/0108207},
 primaryClass = {astro-ph},
       adsurl = {https://ui.adsabs.harvard.edu/abs/2002A&A...381..923S}
}

@ARTICLE{2019ApJ...881L...1F,
       author = {{Fuller}, Jim and {Ma}, Linhao},
        title = "{Most Black Holes Are Born Very Slowly Rotating}",
      journal = {\apjl},
         year = 2019,
        month = aug,
       volume = {881},
       number = {1},
          eid = {L1},
        pages = {L1},
          doi = {10.3847/2041-8213/ab339b},
archivePrefix = {arXiv},
       eprint = {1907.03714},
 primaryClass = {astro-ph.SR},
       adsurl = {https://ui.adsabs.harvard.edu/abs/2019ApJ...881L...1F}
}

@ARTICLE{2008ApJ...679.1422R,
       author = {{Rezzolla}, Luciano and {Dorband}, Ernst Nils and {Reisswig}, Christian and {Diener}, Peter and {Pollney}, Denis and {Schnetter}, Erik and {Szil{\'a}gyi}, B{\'e}la},
        title = "{Spin Diagrams for Equal-Mass Black Hole Binaries with Aligned Spins}",
      journal = {\apj},
         year = 2008,
        month = jun,
       volume = {679},
       number = {2},
        pages = {1422-1426},
          doi = {10.1086/587679},
archivePrefix = {arXiv},
       eprint = {0708.3999},
 primaryClass = {gr-qc},
       adsurl = {https://ui.adsabs.harvard.edu/abs/2008ApJ...679.1422R}
}

@ARTICLE{2009ApJ...704L..40B,
       author = {{Barausse}, Enrico and {Rezzolla}, Luciano},
        title = "{Predicting the Direction of the Final Spin from the Coalescence of Two Black Holes}",
      journal = {\apjl},
         year = 2009,
        month = oct,
       volume = {704},
       number = {1},
        pages = {L40-L44},
          doi = {10.1088/0004-637X/704/1/L40},
archivePrefix = {arXiv},
       eprint = {0904.2577},
 primaryClass = {gr-qc},
       adsurl = {https://ui.adsabs.harvard.edu/abs/2009ApJ...704L..40B}
}

@ARTICLE{2016ApJ...825L..19H,
       author = {{Hofmann}, Fabian and {Barausse}, Enrico and {Rezzolla}, Luciano},
        title = "{The Final Spin from Binary Black Holes in Quasi-circular Orbits}",
      journal = {\apjl},
         year = 2016,
        month = jul,
       volume = {825},
       number = {2},
          eid = {L19},
        pages = {L19},
          doi = {10.3847/2041-8205/825/2/L19},
archivePrefix = {arXiv},
       eprint = {1605.01938},
 primaryClass = {gr-qc},
       adsurl = {https://ui.adsabs.harvard.edu/abs/2016ApJ...825L..19H}
}

@ARTICLE{2017PhRvD..95f4024J,
       author = {{Jim{\'e}nez-Forteza}, Xisco and {Keitel}, David and {Husa}, Sascha and {Hannam}, Mark and {Khan}, Sebastian and {P{\"u}rrer}, Michael},
        title = "{Hierarchical data-driven approach to fitting numerical relativity data for nonprecessing binary black holes with an application to final spin and radiated energy}",
      journal = {\prd},
         year = 2017,
        month = mar,
       volume = {95},
       number = {6},
          eid = {064024},
        pages = {064024},
          doi = {10.1103/PhysRevD.95.064024},
archivePrefix = {arXiv},
       eprint = {1611.00332},
 primaryClass = {gr-qc},
       adsurl = {https://ui.adsabs.harvard.edu/abs/2017PhRvD..95f4024J}
}

@ARTICLE{2006PhRvD..73f1501C,
       author = {{Campanelli}, M. and {Lousto}, C.~O. and {Zlochower}, Y.},
        title = "{Last orbit of binary black holes}",
      journal = {\prd},
         year = 2006,
        month = mar,
       volume = {73},
       number = {6},
          eid = {061501},
        pages = {061501},
          doi = {10.1103/PhysRevD.73.061501},
archivePrefix = {arXiv},
       eprint = {gr-qc/0601091},
 primaryClass = {gr-qc},
       adsurl = {https://ui.adsabs.harvard.edu/abs/2006PhRvD..73f1501C}
}

@ARTICLE{2001ApJ...554..548F,
       author = {{Fryer}, Chris L. and {Kalogera}, Vassiliki},
        title = "{Theoretical Black Hole Mass Distributions}",
      journal = {\apj},
         year = 2001,
        month = jun,
       volume = {554},
       number = {1},
        pages = {548-560},
          doi = {10.1086/321359},
archivePrefix = {arXiv},
       eprint = {astro-ph/9911312},
 primaryClass = {astro-ph},
       adsurl = {https://ui.adsabs.harvard.edu/abs/2001ApJ...554..548F}
}

@ARTICLE{2015MNRAS.451.4086S,
       author = {{Spera}, Mario and {Mapelli}, Michela and {Bressan}, Alessandro},
        title = "{The mass spectrum of compact remnants from the PARSEC stellar evolution tracks}",
      journal = {\mnras},
         year = 2015,
        month = aug,
       volume = {451},
       number = {4},
        pages = {4086-4103},
          doi = {10.1093/mnras/stv1161},
archivePrefix = {arXiv},
       eprint = {1505.05201},
 primaryClass = {astro-ph.SR},
       adsurl = {https://ui.adsabs.harvard.edu/abs/2015MNRAS.451.4086S}
}

@ARTICLE{2012ApJ...749...91F,
       author = {{Fryer}, Chris L. and {Belczynski}, Krzysztof and {Wiktorowicz}, Grzegorz and {Dominik}, Michal and {Kalogera}, Vicky and {Holz}, Daniel E.},
        title = "{Compact Remnant Mass Function: Dependence on the Explosion Mechanism and Metallicity}",
      journal = {\apj},
         year = 2012,
        month = apr,
       volume = {749},
       number = {1},
          eid = {91},
        pages = {91},
          doi = {10.1088/0004-637X/749/1/91},
archivePrefix = {arXiv},
       eprint = {1110.1726},
 primaryClass = {astro-ph.SR},
       adsurl = {https://ui.adsabs.harvard.edu/abs/2012ApJ...749...91F}
}

@ARTICLE{2013ApJ...779...72D,
       author = {{Dominik}, Michal and {Belczynski}, Krzysztof and {Fryer}, Christopher and {Holz}, Daniel E. and {Berti}, Emanuele and {Bulik}, Tomasz and {Mandel}, Ilya and {O'Shaughnessy}, Richard},
        title = "{Double Compact Objects. II. Cosmological Merger Rates}",
      journal = {\apj},
         year = 2013,
        month = dec,
       volume = {779},
       number = {1},
          eid = {72},
        pages = {72},
          doi = {10.1088/0004-637X/779/1/72},
archivePrefix = {arXiv},
       eprint = {1308.1546},
 primaryClass = {astro-ph.HE},
       adsurl = {https://ui.adsabs.harvard.edu/abs/2013ApJ...779...72D}
}

@ARTICLE{2005MNRAS.360..974H,
       author = {{Hobbs}, G. and {Lorimer}, D.~R. and {Lyne}, A.~G. and {Kramer}, M.},
        title = "{A statistical study of 233 pulsar proper motions}",
      journal = {\mnras},
         year = 2005,
        month = jul,
       volume = {360},
       number = {3},
        pages = {974-992},
          doi = {10.1111/j.1365-2966.2005.09087.x},
archivePrefix = {arXiv},
       eprint = {astro-ph/0504584},
 primaryClass = {astro-ph},
       adsurl = {https://ui.adsabs.harvard.edu/abs/2005MNRAS.360..974H}
}

@ARTICLE{2012ARNPS..62..407J,
       author = {{Janka}, Hans-Thomas},
        title = "{Explosion Mechanisms of Core-Collapse Supernovae}",
      journal = {Annual Review of Nuclear and Particle Science},
         year = 2012,
        month = nov,
       volume = {62},
       number = {1},
        pages = {407-451},
          doi = {10.1146/annurev-nucl-102711-094901},
archivePrefix = {arXiv},
       eprint = {1206.2503},
 primaryClass = {astro-ph.SR},
       adsurl = {https://ui.adsabs.harvard.edu/abs/2012ARNPS..62..407J}
}

@ARTICLE{2003ApJ...591..288H,
       author = {{Heger}, A. and {Fryer}, C.~L. and {Woosley}, S.~E. and {Langer}, N. and {Hartmann}, D.~H.},
        title = "{How Massive Single Stars End Their Life}",
      journal = {\apj},
         year = 2003,
        month = jul,
       volume = {591},
       number = {1},
        pages = {288-300},
          doi = {10.1086/375341},
archivePrefix = {arXiv},
       eprint = {astro-ph/0212469},
 primaryClass = {astro-ph},
       adsurl = {https://ui.adsabs.harvard.edu/abs/2003ApJ...591..288H}
}

@ARTICLE{2006ApJ...653L..93B,
       author = {{Baker}, John G. and {Centrella}, Joan and {Choi}, Dae-Il and {Koppitz}, Michael and {van Meter}, James R. and {Miller}, M. Coleman},
        title = "{Getting a Kick Out of Numerical Relativity}",
      journal = {\apjl},
         year = 2006,
        month = dec,
       volume = {653},
       number = {2},
        pages = {L93-L96},
          doi = {10.1086/510448},
archivePrefix = {arXiv},
       eprint = {astro-ph/0603204},
 primaryClass = {astro-ph},
       adsurl = {https://ui.adsabs.harvard.edu/abs/2006ApJ...653L..93B}
}

@ARTICLE{2016PhRvD..93l4066G,
       author = {{Gerosa}, Davide and {Kesden}, Michael},
        title = "{precession: Dynamics of spinning black-hole binaries with python}",
      journal = {\prd},
         year = 2016,
        month = jun,
       volume = {93},
       number = {12},
          eid = {124066},
        pages = {124066},
          doi = {10.1103/PhysRevD.93.124066},
archivePrefix = {arXiv},
       eprint = {1605.01067},
 primaryClass = {astro-ph.HE},
       adsurl = {https://ui.adsabs.harvard.edu/abs/2016PhRvD..93l4066G}
}

@ARTICLE{2007PhRvL..98i1101G,
       author = {{Gonz{\'a}lez}, Jos{\'e} A. and {Sperhake}, Ulrich and {Br{\"u}gmann}, Bernd and {Hannam}, Mark and {Husa}, Sascha},
        title = "{Maximum Kick from Nonspinning Black-Hole Binary Inspiral}",
      journal = {\prl},
         year = 2007,
        month = mar,
       volume = {98},
       number = {9},
          eid = {091101},
        pages = {091101},
          doi = {10.1103/PhysRevLett.98.091101},
archivePrefix = {arXiv},
       eprint = {gr-qc/0610154},
 primaryClass = {gr-qc},
       adsurl = {https://ui.adsabs.harvard.edu/abs/2007PhRvL..98i1101G}
}

@ARTICLE{2007PhRvL..98w1102C,
       author = {{Campanelli}, Manuela and {Lousto}, Carlos O. and {Zlochower}, Yosef and {Merritt}, David},
        title = "{Maximum Gravitational Recoil}",
      journal = {\prl},
         year = 2007,
        month = jun,
       volume = {98},
       number = {23},
          eid = {231102},
        pages = {231102},
          doi = {10.1103/PhysRevLett.98.231102},
archivePrefix = {arXiv},
       eprint = {gr-qc/0702133},
 primaryClass = {gr-qc},
       adsurl = {https://ui.adsabs.harvard.edu/abs/2007PhRvL..98w1102C}
}

@ARTICLE{2012PhRvD..85h4015L,
       author = {{Lousto}, Carlos O. and {Zlochower}, Yosef and {Dotti}, Massimo and {Volonteri}, Marta},
        title = "{Gravitational recoil from accretion-aligned black-hole binaries}",
      journal = {\prd},
         year = 2012,
        month = apr,
       volume = {85},
       number = {8},
          eid = {084015},
        pages = {084015},
          doi = {10.1103/PhysRevD.85.084015},
archivePrefix = {arXiv},
       eprint = {1201.1923},
 primaryClass = {gr-qc},
       adsurl = {https://ui.adsabs.harvard.edu/abs/2012PhRvD..85h4015L}
}

@ARTICLE{2001MNRAS.322..231K,
       author = {{Kroupa}, Pavel},
        title = "{On the variation of the initial mass function}",
      journal = {\mnras},
         year = 2001,
        month = apr,
       volume = {322},
       number = {2},
        pages = {231-246},
          doi = {10.1046/j.1365-8711.2001.04022.x},
archivePrefix = {arXiv},
       eprint = {astro-ph/0009005},
 primaryClass = {astro-ph},
       adsurl = {https://ui.adsabs.harvard.edu/abs/2001MNRAS.322..231K}
}

@ARTICLE{2017MNRAS.472.2422M,
       author = {{Mapelli}, Michela and {Giacobbo}, Nicola and {Ripamonti}, Emanuele and {Spera}, Mario},
        title = "{The cosmic merger rate of stellar black hole binaries from the Illustris simulation}",
      journal = {\mnras},
         year = 2017,
        month = dec,
       volume = {472},
       number = {2},
        pages = {2422-2435},
          doi = {10.1093/mnras/stx2123},
archivePrefix = {arXiv},
       eprint = {1708.05722},
 primaryClass = {astro-ph.GA},
       adsurl = {https://ui.adsabs.harvard.edu/abs/2017MNRAS.472.2422M}
}

\end{document}